# Boltzmann's casino and the unbridgeable chasm in emergence of life research

Elbert Branscomb[*]

Affiliate faculty, Institute for Genomic Biology, UIUC, Champaign-Urbana USA
* brnscmb@illinois.edu

Current Address: 6269 Lariat Loop NE, Bainbridge Island, WA 98110; USA

ORCID
0000-0002-7194-4231

## Abstract

Notwithstanding its long history and compelling motivation, research seeking to explicate the emergence life (EoL) has throughout been a cacophony of unresolved speculation and dispute; absent still any clear convergence or other inarguable evidence of progress. This notwithstanding that it has also produced a rich and varied supply of putatively promising technical advances. Not surprising then the effort being advanced by some to establish a shared basis in fundamental assumptions upon which a more productive community research effort might arise. In this essay however, I press a case in opposition. First, that a chasm divides the rich fauna of contesting EoL models into two conceptually incommensurate classes; here named "chemistry models" (to which class belongs nearly all thinking and work in the field, past and present) and "engine models" (advanced in various more-or-less partial forms by a marginal minority of voices dating from Boltzmann forward). Second, that contemporary non-equilibrium thermodynamics dictates that 'engine-less' (i.e. 'chemistry') models cannot in principle generate non-equilibrium, organized states of matter and are in consequence inherently incapable of prizing life out of inanimate matter.

## 1 Introduction: Boltzmann's order-disorder principle, and the 'cascade economics' of building organized states of matter

A central premise of this paper is that the matter comprising a living system is necessarily in a particular type of extremely organized, far-from-equilibrium thermodynamic state. A circumstance which presents not just a central challenge to comprehending life in general, but one that cannot be validly avoided in speculations as to how 'living forms of matter' could have arisen in the first place. The path to addressing that challenge was opened by Boltzmann through his introduction into thermodynamic theory of "state probability"; and with it the insight that 'ordering' or 'organizing' a body of matter, is no less an act of thermodynamic work than any contemplated in classical thermodynamics. Organizing matter, Boltzmann saw, is putting it in a relatively 'improbable' state and is in consequence *per se* thermodynamically costly; increasingly so the more organized the matter is. But conventional thinking on the problem of abiogenesis (not to mention that in the rest of the biosciences) largely ignores this as a problem, much less as a fundamental law of nature. And when addressing it at all, almost universally gets wrong in what currency, and by what molecular mechanisms, the cost of ordering matter must be paid. The 'unbridgeable chasm' in emergence-of-life theories here being discussed, derives largely from this fact. These are, of course, fairly iconoclastic claims and this introduction seeks to give the reader early warning – and some orientation – concerning the central issues in dispute.

We begin with a brief review of the life and times of Boltzmann's probability epiphany. In the mid 1870's Boltzmann published his most famous and consequential result [1,2]: that the entropy of a physical system is the log measure of the probability of its being in its current "macrostate" (meaning clarified below, see Sec. 6.3). Expressed in terms of the difference in the entropy of two macrostates, that relation is: [1]

$$\Delta S_{a/b} = ln\left(\frac{p_a}{p_b}\right) \quad (1)$$

where the probability of a macrostate is proportional to how richly endowed that macrostate is with microstates (what Boltzmann called "complexions"), again see Sec. 6.3 [2]. Specifically, a macrostate's probability is the ratio of its microstate count to the sum of the microstate counts over all of the system's possible macrostates: $p_a = N_a/\sum_x N_x$ where $N_x$ is the number of microstates in the macrostate $'x'$ – a fact of the greatest conceptual reach and significance as Schrödinger noted in an arrestingly illuminating comment [2] on the fact that Boltzmann's law is fundamentally framed in whole numbers: *"The philosophical implication can scarcely be over-rated. Forces, charges, potentials, collisions – the whole armoury of detailed mechanisms we invent – are apt to retain, in spite of all our striving for the contrary, a slightly mystic tinge or at least much arbitrariness. Boltzmann's discovery, though it cannot entirely dispense with this armoury, assigns to it the rank of an auxiliary device.* **Our true understanding of what happens is reduced to reasoning about whole numbers. We are able to predict the course of events, because we are able to count: one, two, three, four, ... so we can count the eventualities of every thinkable turn and find out the one that is overwhelmingly likely.**"(emphasis added). In this, "every thinkable turn" in "the course of events" is a transition between macrostates, statistically biased to favor those that increase the state's probability – equivalently its count of "eventualities" (i.e. of "microstates") – by the greater factor.

The present piece, of course, is seeking in one or another way to fathom a "course of events"; to make out at least which courses in the emergence of life could in principle have been taken, and which not. And where the governing 'principle', however obscured it may be by the "armory" of "auxiliary devices" whose employment history compels, comes down to simple counting, i.e. to the counting of microstates – and then giving appropriate weight to nature's statistical bias in favor of events that increase that count over those that decrease it. It is thus to the high stools of the microstate counters that we must attend, whether explicitly or not; there seeking especially to understand the paradoxical fact that in certain circumstances nature's impatient appetite for transitions to larger macrostates causes it to 'locally' go against the global imperative, and 'order', or 'organize' a local body of matter into a smaller macrostate (i.e. of reduced microstate count). But it only does so when that move better serves the global imperative; i.e. only when that organization has fashioned the matter into an 'engine' whose operation produces a net gain in the rate at which the microstate count of the total system is increasing; where 'net' means over and above what organizing and operating the engine itself costs. In their basic elements, these facts of nature had Boltzmann's attention right from the start, as we turn to next.

---

[1]This piece adopts the 'physics' convention of treating entropy as a dimensionless quantity; or, to mollify convention, as being expressed in units of $k_B$ – although that convention comes at the high cost of obscuring entropy's "whole numbers" conceptual foundation remarked by Schrödinger (Cf. Introduction). Relatedly, we note that free energy changes (which are properties of a system immersed in a 'bath') are context-dependent surrogates for changes in the entropy of the 'Universe' which is, of course, what the 2nd law constrains (expressed in terms of dimensionless entropies): $0 \leq \Delta S_{Universe} = \Delta S_{system} + \Delta S_{bath} = -\Delta G|_{T,p}/k_B T = -\Delta F|_{T,V}/k_B T$, for, respectively, the Gibbs and Helmholtz free energies (see [3, Ch. 8]).

[2]The physicist Sean Carroll has called the formal expression of Boltzmann's principle, "...one of the most important equations in all of science"; indeed one whose importance "... would be difficult to overestimate." [4, p151,p37]; see also [5, Part 4].

Nearly seven decades after Boltzmann's epiphany, Schrödinger, in his famous 1944 monograph "What is Life?" [6], re-christened that result: "Boltzmann's order-disorder principle". This was to make unmistakably explicit that what Boltzmann's state probabilities measure (as Boltzmann himself very explicitly saw it, see [7] ) is how organized, or ordered, and in that regard how far from equilibrium, the system is. Accordingly, "equilibrium", the most probable of a system's macrostates is also its most disordered, least organized; all others ranking upward in greater order and improbability. A key implication of Boltzmann's insight is that to move a system into a more ordered state its entropy must be reduced. But, of course, systems do not move stably into lower entropy states on their own; and in any case the 2nd law must be satisfied. So how can a system become more organized?

At the accountancy level the inescapable and well recognized answer is that two irreversible (i.e. entropy changing) processes need to be "coupled" together (in effect fused into a single thermodynamic process); one of which is reducing the entropy of the system being ordered, i.e. is "endergonic", the other increasing the entropy, i.e. is "exergonic", of a different system which it is thereby 'disordering' [8]. Of course such couplings are not conventionally described as order-for-order exchanges; but doing so is also making an essential conceptual point as we will see. Beyond that things get harder, and at the same time more central to our concerns. First of all how, in terms of molecular mechanism and thermodynamics, does such coupling take place? Second, how is the process doing the ordering made to proceed in an appropriately targeted manner – to produce, that is, specific organized states with particular properties, not just any equally improbable one? Relatedly, can only certain pairs of processes be coupled in this way? And finally, can such coupling processes themselves arise spontaneously?

Although at the time of Schrödinger's writing the tools needed to address these mechanism questions were not in hand, they were soon in coming; having been already under development for at least two decades based on foundations laid in the 1920's by Théophile De Donder (see: [9, 10]). These developments enabled the principled analysis of irreversible non-equilibrium processes (thereby making truly dynamic "processes" the objects of analytic interest in thermodynamics, not just equilibrium states; and in that distinction an extremely consequential corner was turned). Notwithstanding that limitation, however, by using Boltzmann's principle to cast the two-process coupling requirement in probability terms, Schrödinger brought forward an important conceptual insight, one that is foundational to the present discussion: *that the only way in principle the entropy books can be balanced in the production of order in a system is by the dissipation of order itself; specifically, that of some other system which has 'order to burn'.* Organizing matter, in other words, is not only costly, but the cost can only be paid in the currency of organization itself – by, in particular, *disorganizing* some pre-organized, externally supplied matter. And in the transaction, more organization must be sacrificed than generated.

This 'order-only-from-order' law follows directly from the fact that both of the processes linked in a 'coupling' must satisfy Boltzmann's relation. In probability terms this dictates that at no point in the coupling of a process producing order: ($p_{\text{after}} < p_{\text{before}}$), to one destroying it ($P_{\text{after}} > P_{\text{before}}$), can the joint probability of the two systems be stably reduced; that is, we must have that on average: ($p_\text{after}P_\text{after} \geq p_\text{before}P_\text{before}$). This implies that the exergonic process must be dissipating not less order than the endergonic one is creating (actually more if the coupling is to proceed at finite velocity). And it forces the question around which this entire piece is wound: *How 'in the world' is it possible that disordering one system can (much less alone can) cause the ordering of another?*

In a section revealingly titled: "*Organization maintained by extracting 'order' from the environment*" Schrödinger discusses, in qualitative terms, this key 'order-from-order' issue. The central point being that ordered/improbable states do not and cannot arise or be created 'de-novo'; instead, as we have already noted, it is only by the appropriately coupled sacrifice of order in one system that it can be made to arise in another. This Schrödinger states explicitly when he remarks: "*Thus the device by which an organism maintains itself stationary at a fairly high level of orderliness (= fairly low level of entropy) really consists in continually sucking orderliness from its environment.*" Notice here that it is *not* energy that life must be "continually sucking" in "from its environment" to allow its "orderliness" to be maintained (or for that matter produced in the first place) [3] What must be sucked in is just orderliness itself; just improbability of state!

Arguably, it would be hard to conceive a more profound or important insight than this for understanding life – and its emergence. Yet, although put before us in plain view over 130 years ago it is yet to have a perceptible impact on accepted understandings of how life works, particular among life scientists. This especially so in regard to life's "bioenergetic" mechanisms and, related to that, as to how it might have emerged.

More specifically, conventional understandings in the life sciences recognize neither the critical importance and thermodynamic cost of 'organizing' matter to bring it to life, nor the truth that it is only by the controlled, conditional and 'goal-specific' dissipation of externally supplied ordered states that local matter can be organized and the work of being alive carried out. Much less is it recognized that this ordering of one body of matter by disordering another is not and cannot be carried out simply by mass-action chemistry, however catalyzed or manipulated. Again, this is the primary reason for the 'chasm' being discussed here. Indeed, this question – of how the marked and clearly essential non-equilibrium organization of matter (see Sec. 8.1) in the living state can arise – on which Schrödinger, channeling Boltzmann, so focused – is one which the conventional field of "OoL" research either just sidesteps, or gets wrong. But absent a correct understanding of what it takes to keep the vessels of life organized enough to stay afloat, much less launched in the first place, all further speculation on abiogenesis is, we here claim, futile.

But so too is it absent a plausible path, and a plausible mechanism, for the emergence of organizational complexity itself. A history which progressed, we can with certainty presume, from a minimally organized natal state to the astronomy of organizational complexity all extant life forms display. What forces acting by what mechanisms impose this complexity-favoring developmental bias – are they just thermodynamic, or just biological? And is there a limit to how organizationally complex living matter can become? (See discussions of these points in Sec. 8.1.2 and 12). These questions too hinge on understanding how physical systems can be driven into more ordered, organized, improbable, versions of themselves.

It was noted earlier that Schrödinger and his contemporaries lacked the tools to go beyond the 'accountancy issue', and the implications of the need for process coupling, to meaningfully take on the questions of how such coupling mechanisms actually operate and how they arise. And thus how order is created. This is quite dramatically no longer the case. We now have the tools – including contemporary 'trajectory' thermodynamics [11–13] [4] to ask at the deepest mechanistic levels how such order-creating improbability-for-improbability exchange couplings actually take place.

---

[3] A notion which, in a slightly different context, Schrödinger correctly calls "absurd" – this notwithstanding its being almost universally taken, by lay and eminent thinkers alike, to be a fact enshrined in certainty.

A substantial fraction of the present piece is given to dealing with this specific question of mechanism. And here, we warn, 'there be dragons'; because the 'details', whether they be thought devils or saints, are as counter-intuitive, and as iconoclastic, as things get. The following claims indicate the primary icons set for breakage:

1. "Free energy" is not an energy; much less the 'free part' of one; it is just and only a measure of disequilibrium; specifically how far from equilibrium (equivalently how improbable, or how 'ordered') the system's current state is.

2. Life is not powered by energy consumption. It is powered by dissipating improbability; specifically that of the matter (or photons) imported from the environment in a necessarily improbable state.

3. The coupling of an exergonic to an endergonic process in order to 'drive' the latter does not involve the transfer of energy; specifically not energy 'released' in the driving process being taken up by the driven one. There is no energy transfer in such process couplings.

4. Exergonic-endergonic process couplings are not instances of (catalyzed) mass-action chemistry. They are produced by a task-specific macromolecular device functioning as an 'engine' [5] one which manages individual thermal fluctuations in the two processes such that an anti-entropic fluctuation that produces (by chance) a single-reaction instance in the driven process, is 'trapped' and accumulated after the fact by being made to trigger a single-reaction instance of the driving process.

5. All of the 'improbability power' (aka 'free energy') taken up by a living system is converted in 'order-for-order' exchanges into creating and maintaining the non-equilibrium status of the system.

6. All 'work' performed by living systems, even that seen as purely 'mechanical' (e.g. cargo transport by kinesin) entails moving the material involved in the process into a less probable, more organized (lower entropy) state.

Thus, in exploring the operations of life, one can scarcely place a foot without stepping on one or another aspect of thermodynamic Botzmannian 'improbability'. It is not just that living matter is in a far-from-equilibrium state, nor also that this improbability is produced by a process that 'trades' improbability lost in one body of matter for improbability gained in another. Nor that these trades are effected by exchanging individual and improbable fluctuation events between two processes: an improbability-increasing event in one exchanged for an improbability-decreasing event in the other – which process is thereby relaxing an already improbable state. Nor further, is it just that the ultimate source of improbability driving it all is the extremely 'ordered', improbable, state into which the Universe itself was born [4]). Nor, is it even, finally, that what the living system is doing, thermodynamically, is creating improbably ordered constructions and orchestrations of matter which, paradoxically, enhance the rate at which the Universe is able to disorder itself – in, that is, its ceaseless, statistically-driven pursuit of equilibrium. It is all of these plus the fact that the 'trading' processes which enable it all – which stand at the organizational root of it all – are games of chance governed by the probability rates of individual 'stochastic' molecular events.

---

[4] A closely related approach to the same questions, termed "stochastic thermodynamics" has been developed by several investigators; see, e.g. [14]. Although the differences between the two formulations are not material to our present purposes, in respect of objections raised to the latter [13] we here invoke only the 'trajectory' variant of the contemporary theory of how biological machines, free energy converters most centrally, operate.

[5] A careful reading, preferably before taking on the present paper, of the short 1979 essay by Alan Cottrell: *"The Natural Philosophy of Engines"* [15] is recommended. This in part because it is an exceptionally authoritative and informative – and unique – introduction to the natural phenomena of 'engines'; a concept that is as central to this piece as it is foreign to accepted thinking in the life sciences. Cottrell, (who died in 2012) was an eminent British physicist/theoretical metallurgist. In his essay many of the more important points advanced here receive an illuminating explication. Which, to great benefit, is set primarily in the context of metallurgy, not biochemistry – though with an eye kept fixed on how 'life' works.

Seen rightly, the physical world, life swept along 'downstream' with the rest of it, is a dynamic drama woven, warp and weft, of stochastic fluctuation-driven transitions in Boltzmannian improbability of state – all ultimately seeking to deliver the Universe back to the womb of equilibrium from which it had been, at its birth, 'untimely ripped' [16].

## 2 Background; and the effort to achieve commonality of purpose

> "*All known cosmic and geological conditions and laws of chemistry and thermodynamics allow that complex organic matter could have formed spontaneously on pristine planet Earth about 4,000 mya.*"

The above quotation begins the abstract of a fairly recent review, "*Darwin's warm little pond revisited: from molecules to the origin of life*", by Follmann and Brownson [17]. It succinctly captures the (nearly) universal view that the laws of thermodynamics certainly present no barrier to the assumption that "complex organic matter" can form "spontaneously" (where the text makes it explicit that here 'spontaneously', has its usual meaning in chemistry, i.e. 'proceeds thermodynamically downhill on its own'). And the review goes on to optimistically celebrate *"experimental and theoretical research published over the last two decades, which has added a wealth of new details and helped to close gaps in our previous understanding of this multifaceted field."*

We here make in opposition two claims: first and foremost, that the above thermodynamic assertion is, simply and incontestably, false (as was summarized in the Introduction and will be expanded on at length below). That is, the implied spontaneous self organization of matter by 'chemistry' alone cannot, as a matter of thermodynamic law, take place. Second, that in large part because of this conceptual error, the bulk of OoL research is pursuing directions that could not possibly be fruitful. As a result, the optimistic view of the state of the OoL field is, for the most part, 'off by a sign' – an assessment that is at least not contradicted by the fact that even recent contributions in this field continue to entertain an impressively chaotic and ever-widening mix of conflicting 'origin-of-life' ideas – with no sign of convergence in evidence, or even in sight.

Of course, this is hardly a new characteristic of research on the question – commencing as it did from Darwin's off-the-cuff remark to J.D. Hooker in 1871 [18] – with what is still its most abiding leitmotif; that of "warm little ponds" in which 'chemistry happens' – thereby engendering life [19]. This notwithstanding that the mix has throughout grown greatly in richness and in technical advances deemed "promising". Yet, to the present day, few if any elements of unarguable importance are in hand; and the field struggles to lay claim to real progress. Indeed, all versions and variants of Darwin's warm little pond remain to this day, obdurately sterile.

In implicit acknowledgment of this fact Preiner et al. [20] along with others, have advanced a plea that the field's practitioners seek to establish a body of basic points of agreement on the basis of which their work might become more cooperative, synergistic, and productive. Their effort to this ecumenical end is certainly commendable; in part because it seeks not to find compromise, a recourse not justifiably available to scientists, but simply a common language, a shared conceptual lexicon, and a shared identification of critical questions.

However, I argue here in opposition that as commendable as it is, that effort is also, and where it matters most, futile. That instead, on the bidding-floor of competing ideas, there is an essential divide that cuts to bedrock – arising as it does from irreconcilable foundational preconceptions about the distinguishing properties of living states of matter, about what sorts of thermodynamic realities underpin those differences, and therefore, and especially, about which from nature's kit of mechanisms have in principle the power to engender life. That divide is between two classes of abiogenesis theories [6] here labeled 'chemistry' models and 'engine' models; separated, I argue, by a conceptual chasm far more fundamental and inimical to shared agreements, much less real unifying progress, than is generally thought. And that bridging this chasm is therefore not only infeasible in human terms, it is unattainable on scientific grounds alone; it is just a 'bridge too far'. The main arguments in support of this negative point of view are of two distinct types.

One is framed in the well-recognized terms of early Earth history, geophysics, geochemistry, biochemistry, mineralogy, and standard 'equilibrium' thermodynamics. Michael Russell, Everett Shock, Nick Lane and their colleagues, as well as others, have pressed this facet of the argument in numerous publications [21–25]. The other rests on grounds that are very much less recognized – especially in biochemistry and the life sciences – that of non-equilibrium "trajectory" thermodynamics [11, 26, 27]. But it is also an argument of a different logical character; in the same sense that a technical disagreement about how the electron transport chain functions differs from an argument which, whether recognized to be such or not, is in fact a dispute – as it is in the present case – over whether that system's operation must conform to thermodynamic law. Admittedly, the more dispositive elements of that law are not just sharply unorthodox and counter-intuitive, they have only come clearly to light in recent times; e.g. [28–31]. Furthermore, they are the direct descendants of Boltzmann's probability/entropy insight (cf. Introduction) – whose essential portent has itself been largely ignored by the bio-sciences disciplines. It is not surprising then that violators of this body of thermodynamic law are unaware of the fact.

In any case, this thermodynamic domain of the argument is the subject of the present piece. I turn to it next and begin with a summary of the key thermodynamic issues around which thinking on the nature of life and on its emergence deeply divides.

---

[6] This piece in general abjures the label "origin of life", in favor of "emergence of life". This to honor and reflect that life's genesis is inherently a progressive, incrementally achieved, 'historical' process; one in which each step requires, is necessarily predicated upon, and must be 'vetted' in the context of, its predecessor. The living state of matter is in every valid sense therefor an 'emergent' phenomenon, at its birth and ever thereafter; and is surely destined to remain 'emerging' for as long as it may persist. However, "Origin of Life" is the expression universally adopted by advocates of what we are here calling "chemistry models" reflecting that they in general do not see life as being an instance of a progressive, contingent and inherently historical process of emergence. Whereas models of the alternative "engine" class inherently do. In light of this the two abbreviations "OoL" and "EoL" will be occasionally used to mark and emphasize the non-trivial difference in how the problem under study is conceived in the two classes of abiogenesis models.

# 3 Organizing matter to bring it to life

Our argument is predicated on the claim that the secret of life, both as to how it operates and how it emerged, lies in the far-from-equilibrium organizing of the matter comprising it. Needless to say this assertion is at sharp odds with conventional views; particularly in thinking about life's emergence. In that domain the parting of the ways rests largely on disparate answers to the following questions:

- Is living matter of necessity in an extremely far from equilibrium state – as regards not just its internal 'metabolic' processes but also, and most decisively, its non-random 'organizational' properties? If "yes", is this no less true for life at, or even to enable, its emergence?

- Since a requirement for generating non-equilibrium states dictates that endergonic (entropy reducing) processes must have been active in producing them, how, and why do these processes come into being, and how do they operate consistent with the (post Boltzmann) 2nd law?

- Can suitably catalyzed, purely chemical processes give rise to non-equilibrium states? If so are any of these of possible relevance to the emergence of life?

- Relatedly, in what sense, if any, can non-random organizations of matter arise spontaneously?

As these questions suggest, the pivotal issue in our discussion is that of organizing matter into states which, in thermodynamic terms, are ordered, improbable, and far-from-equilibrium; together with its allied contention that from a thermodynamic perspective generating such states is the essential and core activity not just of metabolism, but of 'being alive' altogether. And here too the 'chemistry' and 'engine' models of abiogenesis are irreconcilably at odds.

## 3.1 The two classes of abiogenesis models

What defines and distinguishes these two classes of models? Much of this essay goes to answering this question, but in summary terms, and in reference to the above specific questions, it is this:

- The core assumption of Chemistry ("OoL") models is that life is fundamentally just suitably catalyzed chemistry; embodied, moreover, in chemical transformations taking place close enough to equilibrium to be conceptually comprehensible in the terms of classical mass-action 'quasi-equilibrium' thermodynamics. And that by implication, to get life launched one needs only to stir up the right 'chemistry': a mix of the right reagents (even often allowing, without noted discomfort, the mix to include various types of pretty fancy "building blocks of life", presumed to have been provided from providence's larder in some manner) in the right concentrations, in the right chemical and physical environment, in the presence of the right catalysts, perhaps being fed "energy" of some type, possibly in the form of certain "energy rich" molecules, or through being excited in a more-or-less chemically non-specific manner into non-equilibrium states, perhaps cyclically driven, etc. – and then wait. At some point a nascent form of life will have 'self-organized' – and from thereon the 'self-organization', just continues autocatalytically! [32–38].

- Engine ("EoL") models, in the sense intended here, posit in contrast that the essential and distinguishing property of matter in the living state is that its comprising atoms are organized into extremely improbable, extremely far from equilibrium, 'ordered' configurations – ordered both physically and in operational orchestration. But of course, matter does not organize itself – does not move itself into persistent lower-entropy, non-equilibrium, improbable states. Instead, as Boltzmann's work brought to light, systems must be thermodynamically driven, and specifically, into each more organized state. Therefore, if to be 'alive' matter must be in a specific type of non-equilibrium, improbable, state of organization, a lead question is how such goal-specific, entropy reducing (i.e. "endergonic") driving is made to happen. And the answer, in a word, is "engines" – that is, by macromolecules that act as engines converting a specific loss of order in one physical system into the creation of specific order in another.

It is worth noting at this point that the 'pro-engine' stance taken in this piece is a direct conceptual metastasis to all order-creating processes of Paul Boyer's "binding-change" model for the operation of ATPsynthase. And that the 'chemistry v. engines' disagreement being discussed here is an exact parallel to the 'Mitchell v. Boyer' battle, in which Mitchell insisted that the operation of ATPsynthase had to be just chemistry, and Boyer countered that a macromolecular engine-like device had to be doing ATPsynthase's essential 'heavy lifting' – namely the conversion of an ion disequilibrium into an ATP v. ADP+Pi disequilibrium (see discussion in Sec. 6.4.2).

In any case, it is here, perhaps most centrally, that 'chemistry' and 'engine' understandings of life – and its emergence – part company. The latter view claims that driving up-hill, order-creating processes is *the* problem life faces; defining in fact its all-consuming metabolic preoccupation. And further, that such driving requires the mediation of a particular type of macromolecular device functioning as an "engine" (see Cottrell, *op cit*). An engine, to make matters worse, whose fuel is improbability of state in one body of matter and whose product is improbability of state in another. And one, still worse, forced to operate by playing molecular games of chance run by 'the house' in a world in which it is that or nothing. And even all of that is 'just for openers'. Every extant living system is a blindingly complex consortium of plumbed-together 'engine' processes generating a great 'cooperative' of distinct but interdependent improbable states; a cooperative which could only have arisen step-wise – each step the result of an incremental chance fluctuation; each viable, if it proves to be, only in the context of its predecessor state and only if in that context it meets the test of providing more gain in order-converting efficiency to the whole consortium than it costs to be maintained. Any credible model for the emergence of life, we engineers contend, must countenance on the one hand this reality regarding the nature of 'matter in the living state' and on the other the step-wise ontogeny of its extreme organizational complexity. One navigated by ascending a staircase of incremental, stochastic, contingent, complexity-increasing organizational transitions – each a chance, fluctuation-driven 'invention' which, moreover, had to be made 'on the fly'.

<div style="text-align:center">

Welcome to Boltzmann's Casino
*You can cash out any time you like but you can never leave* [7]

</div>

## 3.2 Benner and co. to the defense of chemistry

In this author's opinion no better or more important discussion from a biochemist's perspective is to be had of the 'two-worlds' conundrum being here addressed than that given in the 2011 paper by Steven Benner, Hyo-Joong Kim, and Matthew Carrigan: *Asphalt, Water, and the Prebiotic Synthesis of Ribose, Ribonucleosides, and RNA* [39] (hence referred to as the "BKC"). To quote from the introduction to that paper:

---

[7]Modified from the lyrics of the 1977 song "Hotel California" by the rock band The Eagles.

> "In chemistry, when free energy is applied to organic matter without Darwinian evolution, the matter devolves to become more and more "asphaltic", as the atoms in the mixture are rearranged to give ever more molecular species" ... whereas ... "In the bio-sphere, when free energy is provided to organic matter that does have access to Darwinian evolution, that matter does not become asphaltic. Instead, "life finds a way" to exploit available raw materials, including atoms and energy, to create more of itself and, over time, better of itself."
>
> ...
>
> "The contrast between these commonplace observations in chemistry versus commonplace observations in biology embodies the paradox that lies at the center of the bio-origins puzzle. Regardless of the organic materials or the kinds of energy present early on Earth, chemists expect that a natural devolution took them away from biology toward asphalt. To escape this asphaltic fate, this devolution must have transited a chemical system that was, somehow, able to sustain Darwinian evolution."
>
> ...
>
> "Thus, our goal in bio-origins is to identify a chemical system "in transit" from $CO_2$, $H_2O$, $N_2$, and other species that were present on Earth that gained Darwinian capabilities just in time to prevent this fate."

The view of the matter being advanced in the present piece agrees that 'chemistry unaided' is doomed to increasingly "devolve" matter supplied free-energy into molecular chaos, until, at 'asphalt' it can devolve no further [8]. But then it disagrees about the "way" life "found" to avoid that asphaltic fate. Claiming that it was not by riding chemistry part way toward asphalt – far enough, but not too far, that it could then have 'somehow' found a way to jump ship, just in time, from straight chemistry into Darwinian evolution (no little drama in that!) and thereby escape 'death by tar'. To which fate – as BKC correctly argue – it would have been unavoidably destined left in the hands of chemistry.

Rather, we contend, it never was left to those hands, even in its earliest moments; and never took a single step down chemistry's slippery slope towards disorderliness, dissipation – and the tar pits (which BKC invoke). In short, natal life never had to jump ship, and never had to gain the aid of an "in transit" system for help it do so. Instead, as is implied in the Introduction of this piece and belabored in its body, life was and had to be born already 'life-like' – indeed organizationally immaculate; 'organized' at birth (in fact as the essential factor in the act of birth, albeit only in an increment of organization, and as a chance fluctuation) and thereby directly onto the path, straight and narrow, leading up and away from asphalt's hopeless mess of disorganization, toward the far-from-equilibrium organizational states which are, we contend, the truly fundamental, enabling, and universal 'commonplaces' of living matter. In other words, it is here claimed that what is essentially underway in abiogenesis, from its initiating step and thereafter, is imposing order on matter; abiotically initially and progressively – until a point is reached at which it is ordered enough that it can further organize itself 'biotically'.

---

[8] Here an important caveat is required; when BKC refer to "supplying" a system with free energy, they implicitly mean free energy supplied in a 'raising all boats', chemically non-specific, manner and in which its uptake, and dissipation, is not converted to, nor made to be contingent on driving, specific endergonic processes. Whereas free energy supplied in certain specific forms (redox or specific chemical potential disequilibria, along with other improbably ordered forms of matter) can be taken up by specific 'order-or-order' conversion engines, and thereby act to move, or maintain, the system away from equilibrium; away, that is, from the darkness of beckoning tar pits towards the bright light of being alive. But not just 'away from' equilibrium, but into an extremely specific disequilibrium. One comprising a very complex, functionally integrated suite of unitary disequilibria, the organizational design of that suite being a large part of the system's improbability. And all powered by the dissipation of the order of imported, highly specific disequilibria.

Although addressing the question of how this can be is a core purpose of this piece, an indication of where the answer lies is provided by considering the replication of an *E. coli* cell (as we do again later in more quantitative detail; see Sec. 8.1).

### 3.3 Organizing an *E. coli* cell

In minimal medium (augmented with glucose and well oxygenated) such a cell can replicate itself in about 110 minutes. Incontestably, this involves moving most of the matter in the new cell an immense thermodynamic distance 'uphill', 'anti-entropically' into an almost indescribably more organized, farther-from-equilibrium, improbable, 'biological', state – having, as a minor part of that organizing, synthesized almost every molecule it needs 'from scratch'. Although the machinery that accomplishes this levitating leap in extant organisms has certainly been fashioned on the anvil of Darwinian evolution, clearly no such evolution, much less any 'natural selection', is going on during the organizing activity assembling the dynamical wonder of an Escherichia coli cell.

Clearly the replicating cell example puts the question of whether it is possible that the same fundamental principles which in life's maturity are at work in organizing the matter of 'biomass' were, at the dawn of life – though then acting abiotically – responsible for moving non-living matter its initiating steps toward biological organization?

We argue here that it is not just possible but that the alternative is improbable to a near certainty. And of course, the immense 'ordering' achieved in constructing and operating the new Coli cell must not have violated Boltzmann's 'order-disorder' principle, and its implications – as outlined in the Introduction. In particular, each of the ordering processes involved must have been tightly and specifically coupled to a disordering process such that the order (i.e. improbability of state) lost (aka entropy gained) in the second is more than the order gained (entropy lost) in the first. And second, that the coupling of the two processes involved must take place by means of, and at the level of, individual thermal fluctuations (e.g. a fluctuation which by chance, advances a kinesin molecule's foot one step, triggers a subsequent fluctuation which achieves the hydrolysis of an individual ATP). And in which, the incremental factor of probability-of-state lost in one instance of the ordering process, is paid for quasi-coincidentally, 'on the spot' and (in general) after the fact, by its being immediately used to trigger an instance of the 'disordering' process.

As was indicated in the Introduction, the theory that supports the correct analysis of processes of this type is non-equilibrium, 'trajectory' thermodynamics (see discussion in Sec. 6 and Appendices A and B); specifically in the application of that theory to the behavior of a particular type of macromolecular structure functioning as an 'engine' (Cottrell, op cit). Such macromolecular devices operate by executing a particular cyclic sequence of stochastic transitions between distinct macrostates of the system (a 'trajectory') by means of which in each cycle the dissipation of an increment of order in one system is made to be contingent on (typically to be gated by) the prior chance anti-entropic creation of an increment of order in a different system (see Sec. 6.4).

This bringing forward again the dividing chasm between 'chemistry' and 'engine' EoL models; namely the latter's insistence that:

- living systems are, and must be, in far-from-equilibrium states of organization – therein confronting the fact that a physical system can only be driven into organized states by the operation of a specific type a macromolecular device functioning as an engine:

- an engine in particular that produces order in one system in the only way nature in principle allows, that is by 'trading it for order dissipated in another, sufficiently ordered system; i.e. such engines are powered by disorganizing an imported pre-ordered physical system and producing, as their work product, a specifically ordered internal system.

- to do this the engine must manage two irreversible (entropy changing) processes in parallel – while forcing them (generally without directional bias) to run in thermodynamic opposition: one being allowed to take a step 'forward' (exergonically) – i.e. disordering its source system – if and only if the other has 'coincidentally' taken a step 'backward' (endergonically) thereby producing an increment of the 'desired' ordered state.

- where here the 'steps' are individual thermodynamic fluctuation (i.e. stochastic, random) events: an anti-entropic, 'uphill' fluctuation in the process producing order paired with a pro-entropic fluctuation event in the process whose order is being dissipated.

- and in which pairing an 'if and only if' ("neither unless both") rule is enforced, in general by the engine detecting when the desired uphill, order-producing fluctuation has occurred and then responding by 'gating' (i.e. permitting) an instance of its paired order-dissipating downhill fluctuation.

- and, of course, the engine will consistently run in a given direction (i.e. as to which process is driven uphill by the other being then allowed to run downhill) only if less order (less 'improbability of state') is produced in the driven process than is dissipated in the driving one.

Thus, assuming that the above 'order-only-from-order', fluctuation-trading claims regarding how organized, non-equilibrium states are – and only can be – produced in nature, are correct, we trust it is clear that 'engineless' chemistry cannot organize matter (into non-equilibrium states), living matter quintessentially.

But before closing this section we note briefly another dilemma faced by the 'chemistry' school that is related to, but distinct from, BKC's asphalt problem: the so-called 'combinatorial' problem [40]. That is, whether or not asphaltic doom awaits those who rely too long on 'free-energy-supplied' chemistry, any chemical procedure, whether of the laboratory or the cosmos, capable of generating molecules of biological interest produces right from the start, a 'combinatorial' mix of molecular species within which the molecules of interest, biochemists generally concede, are both too rare and too dilute to plausibly initiate 'vital' chemistry. Thus, advocates of the chemistry EoL school generally recognize that the combinatorial problem requires a solution; in some way, a prodigal purification (and also concentration) of the wheat of desired molecules from a suffocating chafe of undesired ones must be achieved – one moreover that can operate 'naturally' on its own – if the whole, it is thought, is to engender life. But even if these feats of concentration and purification were achievable, which as we argue below, they in thermodynamic principle are not, it would be an intrinsically pointless achievement. That because this line of reasoning is chasing a phantom. If just getting the right purity and concentration of the right molecules were the essential prerequisite, then any of the better chemical supply houses would be ready and able to fulfil the order off the shelf. But of course it isn't.

And just as we argued that emergent life never put a foot on the road to asphalt, contra BKC, so too, we claim, it never acquired – nor could it have – its critical molecular constituents – even the much-invoked 'building blocks of life' – by 'import' from the environment. [9] Therefore, emerging life never faced the need to purify, and concentrate, its way out of the dilute combinatorial mess that 'chemistry' unavoidably produces. Had it needed to, however, it would have been at that point immediately blocked from going forward. Deprived of order-converting engines, chemistry on its own simply cannot carry out purifying (or selective concentration) processes; for they too are inevitably creating order, i.e. improbable states, and the Boltzmann relation and its implications applies. How it applies receives a somewhat more quantitative treatment later (Sec. 8.4.1).

But nothing loath on this front from chemists. For them, belief in the existence of some sort of spontaneous purification-by-chemistry process that will solve the combinatorial problem undergirds, whether explicitly or not, essentially all chemistry-based work on the origin of life; see e.g. [41, 42]. And considerable effort has, and is being invested, in attempts to see how this could come about [43–45]. Much of this effort has focused on 'wet-dry cycling' assumed to be able to drive an "evolution" of molecular complexity whose ultimate product would be the full, florid molecular complexity of life – automatically bringing along with it, it is implied, all of the rest of the organizational and behavioral complexities of life. A particularly relevant example is a recent paper by the group of Loren Williams titled "Chemical Evolution Reimagined" [46]; in which they demonstrate "combinatorial compression" through such cycling. Interestingly, the paper begins with this sentence: "*Around four billion years ago, prebiotic chemistry established the molecular keystones of biology, paving the path to life (1). Chemical and geological processes on the ancient Earth caused increases in the complexity of organic molecules, leading ultimately to RNA, DNA, protein, polysaccharides, membrane-forming amphipaths and to the roots of biology.*".

What is in opposition being claimed in this piece is that 'increases in the complexity of organic molecules leading ultimately to .... biology' driven simply by the chemical and physical-chemical processes these authors invoke, flatly cannot happen – being in incontestable violation of thermodynamic law; the law, we remind, governing how improbable, organized states of matter can, and cannot, arise. Not to mention that the proposed ontology of life is also rejected by the 'asphaltic fate' arguments of Benner et al. [39] discussed above.

# 4 Is this disagreement about life's central dependence on far-from-equilibrium, ordered states of matter, and on what it takes to produce them, a real and irreconcilable conflict in OoL thinking?

Yes. A truly unbridgeable gap separates these conceptions. Between them there is no scientifically sensible middle ground and the truth of one is fatal to the other. A situation, moreover, with little more to blame for it than alternative understandings of thermodynamic reality. But there it is. And we stand already at the chasm's edge. We hope the material already presented, particularly in the Introduction (but see also Appendix A), makes it clear enough why this is the case. But because the claims made here are as foreign and heterodox as they are we offer next an alternative way of thinking about the putative 'chasm'.

---

[9] It had to build them all on its own, and teach itself how to do so, step-by-incremental-step, starting with 'fixing' carbon from $CO_2$, 'fixing' nitrogen from nitrate, likely condensing pyrophosphate and polyphosphate from phosphate, and a very small number of other abiotic anti-entropic first steps in the biological ordering of matter (See Sec. 12).

## 4.1 The chasm revisited

When we observe the biomass of our paradigmatic E. coli cell assembling 'itself' (Sec. 3.3) what are we watching? The dominant professional view is framed, at least implicitly, on the metaphor of the "metabolic chart": i.e. that what is going on is just a hugely complex interconnected array of enzymatically catalyzed chemical transformations, all running close to equilibrium; under 'genomic' control and management to be sure – though all of that too is just more 'chemistry' – with the whole operation being necessarily driven by a supply of energy. And then it just 'runs', synthesizing and assembling itself. What else could it be? And therefore that the origin of life must have been in essence the same: very much simpler for sure, and with a lot of the needed parts ("building blocks") being supplied in some abiotic fashion from external sources – all forming a sort of 'homunculus' starter metabolism. And implying that it must be possible to recreate the starter metabolism '*in vitro*'– given the right mix of the right starting abiotic molecules and catalysts, access to the right energy sources, and subject to appropriate abiotic manipulations. Notably, it is taken as a certainty, not subject to reasoned dispute, that some such protocol capable of launching abiogenesis is guaranteed to exist. Again, what else could it be?!

Particular attention is drawn here to one key element of this standard view: that the molecular transformations of life operate near equilibrium. This not excluding metabolically essential but intrinsically endergonic processes. Ones, that is, which have to be driven by being 'coupled to' a stronger exergonic partner.

As has been noted, it is universally believed that in such couplings the exergonic process drives its endergonic dependent by transferring energy to it – thereby fusing the two processes into a unitary chemical reaction – one which, moreover, is generally observed to be only weakly exergonic – and indeed often readily reversible; and thus operating 'close to equilibrium'. But to us disciples of (post-Boltzmann) thermodynamic law not a single element of the above 'just so' story is, or could possibly be, correct. And that when viewing the cell's replication, what the eye of correct discernment sees is something as fundamentally different as different can get. Specifically, it sees:

1. A furiously active 'construction' and 'manufacturing' site – engaged in precisely assembling – out of nothing more than a collection of small molecules in solution – an immensely complex and extremely improbable organization (and orchestration) of matter; an ordered state of the same matter now suddenly standing at an immense distance from equilibrium.

2. Ordered states, however, which can only come into being by a process that is causally linked to the dissipation of at least as much order in some other body of matter (the 'order only from order' principle Cf. Eqn: 1).

3. This implying the requirement for a flow into the cell of matter in a sufficiently organized, ordered, improbable state (e.g. glucose + $O_2$), whose controlled, step-wise, disordering (to $CO_2$ + $H_2O$) can then be 'traded in', albeit indirectly, for the order being created to build and operate the new cell.

4. But, as we noted, Boltzmann's probability insight asserts that such order-trading exchanges can only take place by the 'exchange' of individual thermal fluctuations – which must therefore be individually managed under some strict 'exchange rules'.

5. The lead one of these rules is that the two ("irreversible", i.e. entropy-changing) processes, one losing – the other gaining – order, are required to run in a tightly interdependent manner and also in thermodynamic opposition. Specifically, an individual anti-entropic (i.e. order-creating; "endergonic") fluctuation (having occurred by Brownian chance) in one process (e.g. the transit of an ion across a membrane against a gradient), is detected and used to 'gate' (i.e. permit) an individual pro-entropic "exergonic" fluctuation in the other process (e.g. the hydrolysis of an individual ATP molecule).

6. In which trade, the exergonic fluctuation's only role derives from its relative *irreversibility*. That is, if it takes place before its partner's endergonic fluctuation has reverted, and if the factor by which that event increases the state probability of the matter involved in the exergonic process exceeds the factor by which the endergonic event decreased the state probability of the matter being ordered, then the exergonic event will entropically 'pay for' the endergonic, order-creating one – while also allowing it, in principle, to be prevented from reverting, and potentially therefore added to an accumulation of like events. All of this thereby allowing macroscopic increases in how ordered, how far from equilibrium, the target system is.

7. Whereas, none of the above 'fluctuation management' requirements can be met by mass action chemistry, however fancied up. Instead, as Boyer presciently insisted had to be the case in the operation of the ATPsynthase 'free energy converter', the process management needed for order-for-order exchanges requires the mediation of a macroscopically organized structure; in biology a macromolecule, acting as an engine.

8. In any case, the macroscopically ordered states produced directly, by conversion, from the order brought in from the environment, indirectly drive, via a myriad of subordinate order-for-order conversions, the immense organizational work required to construct and maintain the extremely far from equilibrium, organized, improbable form of matter called 'life' – that not least of the new E. coli cell.

9. And by inescapable inference did so, under the same logic of thermodynamic exigency, to drive and enable its emergence 'in the beginning'.

10. And did so for exactly for the same fundamental thermodynamic reason that, e.g. a thermally disequilibrated (and in that respect ordered) atmosphere can give birth to a tornado. That is, such macroscopically ordered vortex flows emerge first as a single-fluctuation-driven incrementally-ordered 'seed' vortex. But only 'germinate' when that seed of order itself acts as a conversion engine which incrementally, and locally, accelerates the dissipation of the atmospheric stress. And if that acceleration happens to have the positive feedback effect of favoring subsequent fluctuations that increase the engine's size, then such dissipative structures can grow great in size and effect.

11. So too, we assert, did the emergence of life begin with abiotic 'conversion' engines fueled by the 'order' of geophysically-generated redox and electrochemical disequilibria, and producing initially, and locally as 'seed' events, increments of order in the form of the non-equilibrium states of fixed carbon, fixed nitrogen, and perhaps polyphosphates vs phosphate.

12. And finally, that these humbly ordered birthing states would have founded the evolution of a stair-case dynasty of order generation; wherein each step was the predicate for the next incremental increase in organization – but a step only taken if it contributes to the dissipation productivity of the whole 'biological' collective of organizing engines more than maintaining its own organization costs.

Thus, at every turn in the attempt to understand the biochemical machinery of life, most centrally the issue of how it is powered, we are compelled to follow Boltzmann's immersion into the frothing surf of thermal fluctuations – events which, in 'small' systems, can produce qualitatively large effects: e.g. causing stochastic sequences of discrete changes of state in systems comprising a single macromolecule interacting with reagent molecules. And there, it is fair say, fluctuation-based thermodynamics sees the macromolecule-mediated processes of life in the same way they see themselves – i.e. from the point of view of an individual macromolecule (e.g. an enzyme) executing a stochastic, fluctuation-driven cyclic sequence of state transitions in order to mediate an individual instance of a molecular transformation. Not, in other words, as a catalyzed mass-action chemical processes where the catalysis happens to be a macromolecule. And it sees that the macromolecules mediating molecular transformations act as engine-like devices both processing, and driven by, individual thermal fluctuation events. But perhaps the most counter intuitive, even ironic, element of this admittedly strange story is the essential role played by 'scofflaw' thermal fluctuations; those that reduce the system's entropy and mock the conceits of thermodynamic history.

In that connection it is worthwhile to distinguish two quite different contexts in which endergonic fluctuations move a system to a state of reduced entropy. In the first, a single such fluctuation produces the needed change in the system's macrostate; as in the fluctuations that lift either the reactants or the products of a chemical reaction to the (same) transition state. There the utility of that improbable state's relaxation bias is harvested immediately in its disproportionate relaxation to either the reactants or products macrostate. Enzymatic catalysis, of course, manages such exploitations of anti-entropic fluctuations to a stunning perfection. In the second, a quasi-stable and macroscopically significant reduction in a system's entropy must be accomplished, one well beyond the reach of a thermal fluctuation of any finite probability. This requires aggregating a large number of single-fluctuation transitions, each moving a system incrementally along the same path away from equilibrium. In this aggregation, the system mediating it must (on average) balance the entropy lost in each such single-fluctuation transition by entropy gained in another physical system. And must do so while also providing for the aggregation itself, the rapid cycling through replications of the process, high reagent and product specificity, the avoidance of 'leaks' and process abortions, and other matters as well. For this, as we have argued, there is nothing for it but to bring in a true engine (not a catalyst), that can trade in real time an entropy gaining fluctuation in the 'other system' for an entropy losing one in the system being driven into a macroscopically organized state, do so cyclically, and accumulate the incremental products of each cycle.

Thus the view presented here about the organization of matter to bring it to life stands in fundamental and irreconcilable contrast to the conventional view of the matter in two foundational regards. First that such organization is necessary, indeed that an extreme degree of it is the essential property of 'matter in the living state'. Second that the coupling of endergonic and exergonic pairs of processes required to generate organization works by trading organization in one process for that in another – which trading requires the mediation of an explicitly non-equilibrium, 'stochastic', fluctuation-driven, meta-process managed by a macromolecular engine executing Markovian 'trajectories' in the space of the system's macrostates. For these reasons, life at its causal and enabling roots quite fundamentally "transcends chemistry" [47], and conventional chemical thermodynamics along with it.

One consequence of the law that order comes only from order is that the processes that organize matter form a top-down pyramidic cascade of order 'conversions'; each ordered state the 'less-ordered' daughter of one above it. This implies, of course, an infinite regression, one for which nothing less, quite literally, than the emergence of the Universe itself has to answer (See e.g. Carroll, S. 2016; *The big picture: on the origins of life, meaning, and the universe itself* [5]). Another is that only a specific type of macromolecular "engine" can effect the required conversions of order. And a third is that models of life's birth lacking such conversion engines, plus the right sources of 'order' to fuel them, are 'stillborn'; inherently unable to raise matter out of equilibrium's death grip into the exquisitely organized and furiously busy molecular industries of being alive. Whereas instead, 'engineers' hold, organizing matter is, and was from the start, what brings it to life, and that thus to understand abiogenesis, not to mention what in nature creates and maintains 'matter in the living state' altogether, we must understand these order-converting engines: both how they work and how they come into existence. (A simple model system which illustrates all of the important elements of how such 'entropy-trading', order-for-order exchanging systems actually work is presented in App B.)

## 5 How the remaining material is organized

The subsequent sections of this piece mainly undertake to explain and justify the claims made thus far and also to address the core questions raised by them:

1. In what sense if any is it provable that living matter is necessarily in a very far from equilibrium state (in contradiction to the very widely held and published consensus view on the matter)?

2. How in detail do order-converting, thermal fluctuation-driven engines work in both mechanistic and thermodynamic terms?

3. Where do the engines themselves come from, particularly at life's birthing, at which point they would had to have either arisen spontaneously or been supplied abiotically. And how could that possibly come about?

4. Is there clear evidence that engines of the needed types can, and in relevant contexts, do arise abiotically?

5. Could the 'engines' required have been "Self-Organizing Dissipative Structures" (SODS) as initially introduced by Prigogine [15, 48] and thus potentially have arisen spontaneously in purely 'chemical' systems?

In addition, a few other questions are also explored. One is whether there is a universal upper limit on how 'ordered', how improbable, a unit of actively metabolizing biomass can become. Relatedly, what might account for such a limit, and what, if anything, might its existence or its numerical value reveal? Along the way, we discuss what is perhaps the most decisive basis for concluding that mass-action chemistry, catalyzed in any way imaginable, cannot be what life relies on to sustain its work-performing processes, either now or at is birth. This is certainly the least expected, least known, and most 'off the wall' of all such reasons – in part because it emerges from a conundrum, seemingly unrelated, and little remarked, regarding biological "free energy" conversion; see the discussion of this point at the end of the section on Brownian motion (Sec. 6.1).

Others have of course, put forward some or all of the points just made. Particular attention in this regard, for example, is directed to the work of Karo Michaelian and his group; most notably their recent and closely relevant article *The Non-Equilibrium Thermodynamics of Natural Selection: From Molecules to the Biosphere* [49] (which came to my notice during the preparation of this manuscript and is not reviewed or otherwise taken into account here; though note the caveat mentioned in [10]. And in Nick

---

[10]One caveat however; Michaelian's analysis of the emergence of life assumes that it was phototrophic – whereas in my opinion, the evidence against this view is overwhelming [50–52].

Lane's essay "Why Are Cells Powered by Proton Gradients?" [47] he makes the point that the engine converting proton gradients into an ATP disequilibrium, operates via Paul Boyer's "binding change" conformation-shifting mechanism [53], and in that way allows cellular bioenergetics to, in Lane's phrase: "transcend chemistry". Also, in a series of seminal papers Charlie Carter (see e.g. [54, 55]) generalizes, with crushing experimental and conceptual force, these basic insights to other free-energy converting systems. All this in general agreement with the claim made here on thermodynamic grounds that all biological free energy conversions necessarily require the mediation of a task-specific macromolecular 'engine' – devices which are necessarily Boyer-esque shape-shifters and equally "transcend chemistry". But how, exactly?

We turn next to this question: namely how, in fundamental mechanistic and thermodynamic terms, order is traded for order; and both widen and narrow the lens of our earlier discussion of what shifting of conceptual gears is required by the transition to post-Boltzmann thermodynamics.

# 6 The leap to post-Boltzmann thermodynamics: *thermal fluctuations, Brownian motion, conformational fluctuations, and how being buffeted about by water molecules is what brings, and brought, matter to life.*

The Introduction and Appendices A and B) present different approaches to introducing the concepts that form the core elements of contemporary non-equilibrium thermodynamics – to such upending implications for understanding life and its emergence. As these discussions indicate, however, the theory in question is wound throughout on the armature of a single central 'non-classical' thermodynamic concept, that of thermal fluctuations. How can we best come to terms with the conceptual primacy and causal centrality of this 'inexcusably' insignificant, 'nothing burger' of a phenomenon; one to which classical thermodynamics was quite happy to turn a blind eye? First of all, modern thermodynamics can fairly be said to have grown directly out of Boltzmann's order-disorder, entropy/probability insight discussed in the Introduction. The founding elements in which development were those fashioned by Boltzmann himself out of the non-classical conceptual ingredients of thermal fluctuations, microstates versus macrostates, macrostate probabilities, stochastic thermal fluctuation transitions between macrostates, trajectories traversing macrostates comprised of such transitions, and the relationship between a system's probability and its entropy. In this, thermal fluctuations in the 'microscopic' state of a system form the conceptual foundation of the whole – but are also the causal foundation of the physical processes being analyzed. Specifically, they are the 'magic key' for understanding how physical systems can become ordered and living systems in particular organize themselves into the specific states needed to come into existence in the first place and operate thereafter.

However, how the foundational primacy of thermal fluctuations came to be appreciated is not just a remarkable example of the 'emergent' nature of scientific progress itself, one proceeding through a long unsteady sequence of sequentially contingent, but entirely unpredictable discoveries. It is story that is also surprisingly rich in relevance to our current subject. The thread through this history, connecting its first dim intimations to its richly revelatory fruits of present time – including, as will be shown, those lying at the 'beating heart' of our present topic – is "Brownian motion".

## 6.1 Brownian motion

In 1828, the eminent Scottish botanist Robert Brown (F.R.S. Hon. M.R.S.E. & R.I. Acad. V.P.L.S.), published a paper titled "*A brief account of microscopical observations made in the months of June, July and August 1827, on the particles contained in the pollen of plants; and on the general existence of active molecules in organic and inorganic bodies*" [56] – followed the next year by an addendum "*Additional remarks on active molecules*" [57]. In these papers Brown reported his detailed studies of the observation that particles ("molecules") of a few microns in size suspended in water moved incessantly in an "agitated" fashion. While initially seen with particles extracted from grains of pollen, Brown painstakingly extended these observations to particles of like size from many different sources. Certainly a puzzling observation at the time, and in fact one that had been made, if less systematically, by many since the invention of the microscope [58].

But here we must not just slide by the fact that Brownian motion is something that particles "suspended in water" show. Since absolutely central to our story is the fact that the macromolecules of life are "suspended in water" and are of a size for which Brownian motion is a dramatic reality – and the beckoning door to a deep epiphany – namely how the 'organized' states of living matter come into existence, and operate.

All this the fruit of Dr. Brown's "minute" observations, notwithstanding that to modern eyes it would be hard to imagine a humbler series of scientific investigations than Brown's and the work is a moving testimony to the correspondingly humble state of knowledge in the early 1800's. Yet it set seed to one of the most consequential flowerings of scientific understanding in history – a seed, however, as the 'progress' of science so often sees to it, that was in no apparent hurry to germinate. In Bertrand Duplantier's rich monograph on the history of Brownian motion [58] he notes that "*Between 1831 and 1857 it seems that one can no longer find references to Brown's observations, but from the 1860's forward his work began to draw large interest.*". Nor even thereafter however was progress any steadier, to the honor of none of the relevant disciplines. Of this early part of the history, the Nobel laureate Jean Perrin, to whose critical contribution to the story we return below, noted in 1909 that Brown's observations, and the phenomenon altogether, were "*for a long time ignored by the majority of physicists*", leaving it to just a few who "*recognized in this, supposed insignificant, phenomenon a fundamental property of matter*" [59]. But a very fundamental property of matter it has indeed proven to be; and not least to understanding the workings, and the emergence, of life. A defining break in this protracted neglect came in 1888 through the publication of a remarkable study and analysis of Brownian motion by M. Gouy [60] which established (to again quote Perrin (Op-cit)) " ... *not only that the hypothesis of molecular agitation gave an admissible explanation of the Brownian movement, but that no other cause of the movement could be imagined, which especially increased the significance of the hypothesis. This work immediately evoked a considerable response, and it is only from this time that the Brownian movement took a place among the important problems of general physics.*". A place of importance however which still did not produce its essential epiphany until 1905, nearly two decades later, when Einstein published a convincing theoretical explanation of the phenomenon [61] [11]. But the major import of this work was not that it explained Brownian motion, but how. By adopting Boltzmann's atomistically discretized view of reality, and making use of the Maxwell-Boltzmann velocity distribution, Einstein showed quantitatively that, just as Gouy and others had earlier conjectured, the Brownian motion of small particles in water reflected statistical fluctuations in the thermally excited impact of water molecules on the particles [12]. Yet, such is the high

---
[11] A result coincidentally and independently obtained by Sutherland; see discussion in [58]

[12] In particular, Einstein analyzed Brownian motion as a 'random walk' diffusion process giving rise to

static friction against which science moves, it was still not until the publication in 1909 by J.B. Perrin of decisive quantitative experimental studies confirming Einstein's results that their conceptual content gained general acceptance as actually reflecting reality [59]! An acceptance which then played a key role in finally hammering the coffin shut on opposition to the existence of atoms [63, Ch. 11]. In recognition of which fact Perrin was awarded the Nobel in 1926 for having established the reality of atoms [64]!

But it was not just the reality of atoms that all this work had now forced on the scientific mind, it was in addition and no less fundamentally, the centrality of thermal fluctuations in determining the properties and the behavior, of all manner of 'soft matter' phenomena [65]; and not least of that matter whose particular softness brings it to life. An intellectual awakening whose superordinate importance is reflected in noteworthy assertions of parentage. In expanding on Einstein's role in it, Frey and Kroy remark, for example, "*Without exaggerating, he may therefore be called the father of fluctuation theory*" [65, p4]. Perhaps. But if Einstein was father to the 'fluctuation' insight, Boltzmann was grandfather.

And in the now fairly settled dust of what they together begat, a valid idea about the fundamental nature of life not wound on the armature of 'thermal fluctuations' can hardly be framed. Indeed, and not to slight electrons, Albert Szent-Györgyi, could arguably have cut to a deeper truth had he quipped, "*life is just some thermal fluctuations 'breaking bad', so that others might with greater haste 'break good'*". This is the case, it turns out, because it is by exploiting specific Brownian fluctuations occurring in macromolecular structures of appropriate types that nature trades order-creating thermal fluctuations in one system for order-dissipating ones in another. But to understand this connection between Brownian motion and the generation of life's organized states it is useful to follow further the particular thread of the Brownian motion story that has, especially more recently, led to insights, both critical and counter-intuitive, into specifically how life's molecular machinery operates.

But that part of the history is also one of fits and starts and another half century passed after Perrin's work before it began to be realized that Brownian motions were not just hapless witnesses to the reality of atoms, but were, among other things, an essential, life-enabling aspect of the behavior of macromolecules in water. This reflecting that it was then being seen that the function of all of the essential machines of life: enzymes, free-energy converters, and more, depended on conformational fluctuations within or between domains of these macromolecules and that these conformational fluctuations were instances of Brownian motion (see [54, 62, 66–70]). We note in particular Astumian's remark that "... all chemically driven molecular machines in water, ..., are properly termed "Brownian Motors" [71, p1720]).

Then another few decades passed before our story's *coup de grâce* realization emerged, this time at the hands of Dean Astumian and his colleagues ( [12, 72, 73]). Namely that because these functionally essential conformational fluctuations involve protein-sized objects executing necessarily over-damped motions through water (harkening back to Einstein; recall Sec: 6.1) they take place at the so-called 'terminal velocity'; a speed which is proportional to the thermodynamic force directionally biasing the fluctuations. This directly has the quite startling further implication that the free energy conversions of life necessarily satisfy Onsager's 1930's assumption of flux-force linearity, and therefore his so-called reciprocal relations *at all driving strengths*, not just, as he and all others 'knew' the realities of physical chemistry demanded, for those in the tepid shadows of equilibrium [62]!

And this final fact, it turns out, is a true, quite literally 'life saving' *'deus ex*

---

what is now called the "Stokes-Einstein" equation; this expresses the random walk's diffusion coefficient $D$ in terms of the viscosity of water $\zeta$: $D = k_B T/\zeta$. We here note for later the essential involvement in Brownian motion of the finite viscosity of water. And also note here, and explain later, the quite stunning inference that were this not the case, matter could not have come alive; see discussion in [62].

*machina'*. In light, that is, of the unchallenged conclusions reached in an earlier (1983) effort by Jörge Stucki and colleagues to understand the observed, but at that time entirely unexplained and 'law-violating', flux-force linearity of the free energy conversions carried out in living systems [62, 74]. These conclusions are two-fold. First, the demonstration that were life's conversions operating under the exponential flux-force dependence dictated by standard theory (e.g. Hill [75]), they would be "at least" millions of fold less efficient than they in fact are. Second, a compelling argument that metabolism would be entirely incapable of 'making a living' were it working with such extremely low conversion efficiencies.

**The great and vital 'efficiency' gift of Dr. Brown's thermally-driven 'agitations'**

So for this most indirect and unanticipated reason, the fact that the operation of life's macromolecular devices depends on Brownian conformational fluctuations taking place in water is what makes it possible for life's metabolism, its free-energy conversions most directly and essentially, to 'affordably' organize matter as needed to bring it to life – and hold it there!

What a strangely consequential 'pinball' journey, of almost 200 years, is this; especially one initiated by so humble a crumb of evidence as that brought to light by Dr. Brown's "minute" microscopic observations: a journey into, as it became evident, the hidden effects of thermal fluctuation-driven impacts of water molecules on small particles. And taking us all the way from the strange, unexplained, seemingly 'life-like' agitations of particles in aqueous suspension glimpsed by Brown, to the highly specific and orchestrated movements, also in water, of the domains of life's macromolecules, those of its 'engines' most essentially – without whose dynamic agency, it is now possible to conclude, matter would have remained ineluctably lifeless.

Obviously, these points regarding Brownian motion and thermal fluctuations, 'retrograde' ones especially, add force to the view advanced here in opposition to speculations on the emergence of life not founded on the need for, and the centrality of, these engine-based, fluctuation-driven thermodynamic principles regarding how material systems are organized. But we are now at the point where we can address in an appropriate context the mechanistic question of how 'exactly' it is that thermal fluctuations driving Brownian motions within and between macromolecular moieties, permit order to be gained by being traded, bit by bit, for order lost.

## 6.2 Boltzmann's casino and thermal fluctuations that decrease a system's entropy, 'violate' the (pre-Boltzmann) 2nd law, and organize nature

Standing back for the moment from the 'order-for-order' claim made above, the base-level question undergirding the present discussion is this: what in principle can force a system to move at macroscopic scales, and progressively, away from equilibrium into a state that is more ordered, more improbable, and therefore of lower entropy? In an importantly revealing but also delightful sense, the answer is 'nothing'. This is because, as we noted above, the only event in nature that can move a system entropically up hill is a chance anti-entropic thermal fluctuation in that system; those events Boltzmann raised from the banishment of outlawry demanded by orthodox theory into the light of reality, there endowed moreover with the premier status of causal essentiality.

However, as we also noted above, such fluctuations are the exclusive province of the eponymous Dr. Brown; who deals them out by a fair sampling from, typically, a Gaussian distribution. "Fair" in that he can neither be commanded nor influenced as to what fluctuations he should pull out of his Gaussian hat, or when he should do so. Nature is obliged therefore to be a true and unyieldingly patient Zen archer; and simply wait until the good Doctor happens by pure blind chance to offer up one of the extremely rare specific fluctuations that can move any particular system some distance toward a specific more organized state.

But here nature faces a problem; actually an impressively daunting bunch of them: One we just noted: that thermal fluctuations are random, both as to time and direction. Also, the desired 'retrograde' order-generating ones individually produce increments of progress too small to meet the need. Even more problematically they are also inherently transient, reversing rapidly and with a probability rate that grows exponentially with the size of the desired fluctuation (See Eqn. 2). Also, as has been noted, the occurrence of a specific fluctuation that could contribute to a given order-generating progression is itself a very rare, improbable event – so in general almost all of the fluctuations being offered must be stringently ignored. Finally, the fact that each of the thermal fluctuations of interest reduces the entropy of the system only incrementally implies that a large number of like macrostate transitions must somehow be serially accumulated. This in turn implies that for one such fluctuation-driven change to be of any use, it must be both 'captured' and 'payed for' thermodynamically before it reverses, that is 'on the fly'; only then can it form the macrostate predicate for a subsequent incremental ordering step of the same type – while further standing ready to do so recursively.

All this then allowing the system to ascend, one providential fluctuation step at a time, a staircase of incremental elevations in improbability toward greater order. Not to mention, especially in the edifices of life, ones of pure magnificence. To put this in other terms, retrograde thermal fluctuations are quite literally the only material nature has to work with to produce macroscopically organized non-equilibrium states of matter. And if it is to get anywhere in the building trades it simply has to find a way to assemble a 'be-spoke', precisely ordered structure, one brick at a time, out of these tiny, fly-by-night, 'wrong-way', random thermal fluctuation events. Seems a hardly sensible, not to say credible, way to go about things. But of course, we can but play the ball of reality where it lies. And that turns out to be in part where Boltzmann put it when, as the man who famously reified atoms [63, 76]), he did the same for '2nd law-violating' thermal fluctuations. Though in so doing he also laid the essential foundations for many important subsequent advances. Ones that, among other things, exposed the ironically paradoxical role played in the affairs of the Universe by these long-neglected scofflaw fluctuations. Namely that they are the causal source of all work, of all organization, of every 'improbable' non-equilibrium state of matter, even of every chemical reaction's surmounting of a transition state. In consequence, of everything in nature that is wondrous and worthy of our study. Entropy-decreasing fluctuations, it is indeed strange to contemplate, are the fabric of creation and the fountain of all order.

Still to be directly addressed, however, is how this mad fluctuations-producing-order business actually works down in the molecular engine room. In that story, as will become plain, the crux move is the final point listed in the above litany about how order is traded for order: the need to capture and thermodynamically pay for individual retrograde fluctuations in real time when they occur by chance. It is at this point, of course, that the law adduced earlier, that only order in one system, through its sacrifice, can give rise to order in another, must enter the story. But how? How, out of the froth of thermal fluctuations is the dissipation of a bit of improbability of state in one physical system to be traded for the stable and persistent creation of a like increment of improbability in another? We have presented the answer to this question qualitatively in the above, but turn now to doing so in more formal terms. And for this, we need wade a bit deeper into post-Boltzmann thermodynamics.

## 6.3 Microstates, macrostates, and the stochastic trajectories between macrostates that achieve conversions of order

This is again a story of thermal fluctuations; specifically those responsible for the fact that a physical system is incessantly moving from one 'microstate' – specified by the values of its "degrees of freedom" – to another. A story which begins with noting the essential distinction between changes of microstate that make no 'sensible' difference in the system, and those, very much rarer ones, that do; and ends by clarifying the paradoxical sense in which it is the differential abundance of microstates, across "macrostates" (defined immediately below), that determines the dynamic behavior of the latter. Stated in somewhat different language: due to thermal fluctuations a physical system moves unceasingly between different 'microstates'. Almost all of these transitions, however, are 'invisible' in that they change no observable property; only rarely does such a fluctuation lead to an observable, i.e. "macroscopic", change in the system. The lead implication of this fact is that a system's microstates can be exhaustively partitioned into non-overlapping subsets termed 'macrostates', the members of each of which, by definition, differ in no measurable property. Thus each possible specification of the values of system's observable properties defines a different macrostate, and each microstate is in one and only one macrostate.

The first thing to note about this partitioning of microstates into macrostates is that it divides up the set of microstates extremely unequally. This is a fact of great consequence since pretty close to 'everything' about both the static and dynamic properties of a system is determined by the differences in the 'richness' with which its macrostates are endowed with microstates. It was in fact just this difference that led Boltzmann to his 'entropy is probability' insight [13]. Second, since the system has an equal probability of being in any one of its microstates, the probability of it being in a particular macrostate is simply proportional to the number of microstates it comprises (often referred to as the macrostate's 'multiplicity'). This fact is what underlies the remark by Schrödinger noted earlier regarding the ascendancy of simple 'counting' in Boltzmann's great reconceptualization of thermodynamic theory (quoted in the Introduction), i.e.: *...We are able to predict the course of events, because we are able to count ...*". And it is a fact of such transcendent significance that it stands alone on Boltzmann's gravestone (albeit in the regrettably bowdlerized form: $S = k \log W$, rather than in the correctly, and authentically Boltzmann, dimensionless form $S = \log W$).

Of course, beneath these static properties a boiling froth of thermal fluctuations is keeping the system in constant motion – albeit with observable consequences only for those relatively rare fluctuations that transition the system to a different macrostate. Furthermore, the microstate-changing fluctuations that move the system to a new macrostate are blind as to their destination, and so are more likely to land on a microstate in a larger, than in a smaller, macrostate. Thus the path of a system through macrostates is a stochastic, directionally-biased, 'pin-ball' decent through a sequence of states trending to ever larger ones. This underlies the classical version of the 2nd law's assertion that the 'goal' of all natural processes is to increase the entropy of the system in which they are active until the macrostate of maximum entropy, called "equilibrium", is reached. (see [7, §89]). That is, a system's observable dynamics takes the form of a stochastic "trajectory" comprising a linked sequence of time intervals wherein each of which the system is 'wondering around' within a given macrostate, punctuated by relatively instantaneous transitions to a different macrostate; and defining a sequence which therefore has the character of a biased, (Markovian) random walk. Importantly, the transitions in the sequence are stochastic; random as to both time and direction, including in particular having a finite chance of reversing any transition. This brings the differential microstate richness of macrostates forward as the key to a system's fluctuation-driven dynamical behavior. A fact fairly recently given quantitative expression through the discovery of so-called 'fluctuation relations' [30, 31, 77–79].

---
[13]Though not directly; in between stood his work on gas theory, and the so-called "H" theorem; see [7, Ch. VII].

About which we merely note that in systems of the type considered here, which involve trajectories of "diffusive" transitions between macrostates whose probabilities are unchanging in time, the applicable diffusion relation takes a particularly pleasing and simple form (see Equations 2 and 12 and Secs III and IV) in the 1999 paper by Gavin Crooks: *"Entropy production fluctuation theorem and the nonequilibrium work relation for free energy differences* [30]). In particular Eqn 2, from that paper is:

$$\frac{P_F(w)}{P_R(-w)} = e^{(+w)} \tag{2}$$

*"where w is the entropy production of the driven system measured over some time interval, $P_F(w)$ is the probability distribution of this entropy production, and $P_R(w)$ is the probability distribution of the entropy production when the system is driven in a time-reversed manner."*

That is (simplifying a bit), transitions to states of greater 'size', i.e. greater probability – greater entropy – are, in the magnitude of the entropy gain, exponentially more likely than the reverse, larger-to-smaller transition.

On this intro to 'fluctuation' thermodynamics, we can now delve into those special trajectories that trade order for order, and thereby do the work of organizing matter.

### 6.4 Trajectories that trade order supplied for order created

As we noted above, these are cyclic trajectories executed by systems that are 'coupling' two irreversible processes taking place in different systems in a way that forces them to move in thermodynamically opposed single fluctuation steps – such that if by chance one fluctuates "up" (endergonically, i.e. 'backwards', and thus produces an increment of order) the other must, in the same cycle, then fluctuate "down" (i.e. 'forward', moving exergonically, and destroying order) – or the cycle has to start again. And as we have emphasized, imposing the necessary conditional and intertwining order on the fluctuation transitions executed by the coupled partners requires the engagement of a macromolecular machine acting as an engine.

However, it has more recently come to be understood, as Astumian has in particular emphasized [80], but see also Seifert [14]), that a single macromolecule in solution interacting with small molecule reagents has in general a sufficient number of 'mechanical' degrees of freedom, and spends sufficient time in any one of its macrostates for each such macrostate to reach equilibrium with respect to those degrees of freedom – even though the system will in general not be in equilibrium with respect to the chemical (or electro-chemical) potentials involved – and will therefore not be in the system's equilibrium macrostate. Each macrostate in such a system is therefore, meta-stably, its own 'classical' mechanically equilibrated thermodynamic system; i.e. with its own well-defined internal energy, entropy, and Helmholtz free energy ( [14, 81]). Trajectories in such systems are paths of transitions between these 'classical' quasi-equilibrium macrostates of the system, paths acting to relax the out of equilibrium state of the system with respect to its chemical degrees of freedom.

#### 6.4.1 An example order-converting trajectory

Consider that we have in solution four reagents $R$, $P$, $S$, $Q$ supporting two explicitly reversible reaction processes, $R \rightleftharpoons P$ and $S \rightleftharpoons Q$; where it is arbitrarily assumed that in these their chemical potentials are such that $R \to P$ and $S \to Q$ are the reaction's forward, exergonic directions. To this is added a single "engine" macromolecule "$E$" capable of executing a specific kind of cyclic trajectory in which it manages pairs of individual events from the two reactions. In this it enforces the following rule: an individual cycle is allowed to complete only if in it a single *endergonic* event from (either) one of the processes is paired with a single *exergonic* event from the other. It being understood that since these are all single-molecule events they are necessarily manifestations of individual thermal fluctuations – and that we are dealing therefore with an explicitly stochastic, probabilistic process.

How is this sort of dual-reaction process made to happen and why exactly does it do the job? The sketch overview is that the two reaction events are each split by the engine into half reactions which are then interleaved, again by the engine, in a particular sequence of 'contingent' fluctuation-driven transitions (as originally proposed by William Jencks [82]). In the present model system, these are in order: i: bind Q, ii: bind R, iii: catalyze $S \leftarrow Q$ (the key endergonic fluctuation) and release $S$, iv: catalyze $R \to P$ and release $P$ – and then return to 'go'. Here "iii" is the work performing, anti-entropic fluctuation, and "iv" the pro-entropic 'latch' fluctuation, triggered by "iii", which traps (after the fact!) and preserves the work fluctuation. This, of course, is not describing a mass-action solution chemistry process, or even just a collection of catalyzed reactions. It is a disciplined 'trading' of individual thermal, thermodynamically opposed, fluctuation events one at a time and in a particular order. And, in the conversions of biology, each such trading process is carried out by a highly specific, purpose-built, macromolecular engine. As, we argue, they have to be; because on fundamental thermodynamic principle this is the only way 'order' can be imposed on molecular systems. Note also that, just as is the case for the majority of life's conversions, there is no necessary 'chemical' connection between the two processes being coupled; any pair of processes over which a macromolecular device could in principle enforce the required interleaved half-reactions contingency rules can be used to cause order gained in one to be traded for order lost in the other.

To put this in more explicit (Boltzmannian) thermodynamic terms, the engine, together with the reagents, defines a set of distinct macrostates (recalling the "Binding Change" mechanism introduced by Paul Boyer to explain the F1 engine of ATPsynthase). In our simple example these are as given in Table 1. Importantly it is further presumed here that in each of these macrostates the engine is in a distinct conformational (allosteric) state (indicated by superscripts), each having a distinct binding or catalytic capability as indicated in the note for that state. In consequence, a transition to a new macrostate in the cycle involves in general changing both the system's chemical state and the engine's conformation (see expansion on this important, and somewhat 'non-standard', element of our version of conversion engines below in Section 6.4.2).

The following points about such order-creating cyclic trajectories require emphasis.

1. In the rendering of the cycle given in Table 1, transitions between its macrostates involves a chemical event, followed by a conformation change in the engine triggered by that event. And that it is the conformation change that in turn enables (and thus 'triggers') the subsequent state to carry out its own specific chemical transformation (but see comment on this point following this list and also Sec. 6.4.2).

2. If the only permitted transitions are between neighboring states in the order they appear in the table, the system operates as an ideal converter, in which the only 'wasted' moves are those fluctuations which move the system one step back up the sequence. But real-world systems experience other, 'off-sequence' transitions between its macrostates, leading to unproductive cycles and degraded performance.

**Table 1.** Macrostates, and transitions between them, as imposed by the engine $'E'$ for a simple order-converting trajectory in which the reactions $R \rightleftharpoons P$ and $S \rightleftharpoons Q$ are 'coupled' in thermodynamic opposition. The job of the engine is to insure, as far as possible, that a cycle completes if and only if a single instance of the two reactions, one preceding endergonically, the other exergonically, has taken place. (In the table the symbol $'[rX]'$ stands for the four reagents $R, P, S, Q$ free in solution and their molecular numbers at the start of the cyclic conversion trajectory), where the cycle preceding in the top-to-bottom order produces one step in the order-creating, endergonic process $S \leftarrow Q$ at the expense of one step in the order-dissipating, exergonic process $R \rightarrow P$. Note: to simplify the table, the cycle's two 'reactions', steps #2 and #3, combine what are in fact two separate steps: the reaction itself and the release of its product.

| State Indx | Macrostate | Note |
|---|---|---|
| 0 | $[rX]\,E^0$ | $E^0$ can bind Q, and then change conf. to $E^1$ |
| 1 | $[rX - 1Q]\,QE^1$ | $QE^1$ can bind R; $E^1 \rightarrow E^2$ |
| 2 | $[rX - 1Q - 1R]\,RQE^2\text{x}$ | $RQE^2$ can cat $S \leftarrow Q$; then release S; then $E^2 \rightarrow E^3$ |
| 3 | $[rX - 1Q - 1R + 1S]\,RE^3$ | $RE^3$ can cat $R \rightarrow P$ release P; then $E^3 \rightarrow E^4$ |
| 4 | $[rX - 1Q - 1R + 1S + 1P]\,E^4$ | $E^4$ can relax to $E^0$ |
| 5 | $[rX - 1Q - 1R + 1S + 1P]\,E^0$ | back at 'start'; one cycle completed |

A rigorous treatment of conversion systems must therefore give each of these alternative trajectory paths its correct statistical weight (see e.g. [75]). With that in mind, however, we can confine our attention to the ideal converting pathway.

3. An individual transition between two macrostates is necessarily a transition between individual microstates in those macrostates, and therefor involves a random sampling from an immense set of alternatives sequences of microstate transitions – involving all choices of specific starting and ending microstates and all possible choices of when in time each transition takes place and whether it is moving forward or backward in the sequence. In formal terms, the challenge is to determine the probability distribution over the set of microstate-mediated trajectories passing through the specified macrostates. The foundation of the theory of such trajectories therefore rests on defining the probability distribution over all legal individual transits [11, 75].

4. In this description of the engine's operation, being based on stochastic trajectories in which the transitions are controlled by probability rates, it is implicit that the operation of these engine systems is inherently kinetic with built-in kinetic 'competitions' between alternative paths. In this, one such is of particular note; that reflecting the high desirability, on efficiency and resource-sparing grounds, of trapping a favorable endergonic fluctuation (the occurrence of the 'back reaction' fluctuation: $S \leftarrow Q$ which can occur in state #2). If it does, the system can move to 'trap' it, first by transitioning to state #3 in which it can catalyse the exergonic 'latch' reaction: $R \rightarrow$ P – and release of P; and second by following that reaction by moving on to states #4 and then #5 to complete the cycle. Basically, once the latch reaction has happened, it would have to be reversed before the endergonic one could also be reversed. The key 'race' competition is then between the reversion of the endergonic reaction, and the initiation of the exergonic 'latch'.

Note that (ref. Boltzmann's order-disorder relation; Eqn. 1), if the irreversibility (the probability-gain factor) of the $R \rightarrow P$ reaction is greater than the probability loss factor in the $S \leftarrow Q$ order-creating 'back fluctuation', then the system not only has a better than even chance of winning the kinetic race in the desired direction, but at the same stroke 'incidentally' paying back with profit the entropy lost in the order-producing fluctuation.

5. Note finally, that the only role played by the 'driving' process – quintessentially,

the hydrolysis of ATP – is to come in after the fact and, through its relative irreversibility, act as a latch to trap (if it has the good statistical fortune to do so), the thermodynamically 'uphill' Brownian fluctuation which, entirely by chance, produced one precious, small, increment of the 'disequilibrium' (i.e. the 'order') the engine is working to create. And again, no 'energy' changes hands.

Clearly, the process by which order is traded for order ("free energy is converted") isn't just a chemical one; instead the most central process in bringing matter to life, the harvesting of free energy (aka 'order') requires a (macromolecular) engine. But we have also introduced an element into the story of such conversions that is not a standard or settled consensus matter. That having to do with the role, here assumed to be ubiquitous and essential, of conformational changes in mediating the macrostate transitions in the conversion engines of life. This is taken up next.

### 6.4.2 Mitchell v. Boyer, chemistry v. machinery – in a greatly expanded venue

The assumption made here that the macrostate transitions in a free energy conversion trajectory involve rate-limiting conformation changes has been advanced by other investigators; commencing in 1976 [66] but aggressively advanced more recently by Dean Astumian [12, 83–87] although usually in reference to specific conversions, e.g.: [88, 89].

It must be acknowledged, however, that this assumption is at consequential odds with the standard treatment of free energy conversion (as embodied, e.g., in the field's bible by T.L. Hill [75]). In that treatment, all of the transitions of the cycle – save, typically, a purely conformational switch between e.g., states 1 and 2 in Table 1, and then, its reverse between 3 and 4 – are seen as arising from purely chemical changes (binding, unbinding, catalysis) and the conformation shifts are moreover taken to be instantaneous. These assumptions have the implication that the chemical changes are the rate limiting processes in the conversion cycle, and therefore that the "flux-force" dependency of conversion rates on the free energy changes driving them must be exponential.

The problem, arrestingly, is that reality 'didn't get the memo'; the free energy conversions of biology are, when appropriately characterized, uniformly observed to deviate dramatically from this prediction and to instead obey, quite steadfastly, linear flux-force relationships (see discussion in [90]); this including, most stunningly and dispositively, the entire OxPhos chain and all of its constituent conversions [74]!

In the conclusion to our earlier discussion on Brownian motion (Sec. 6.1) we explained what underlies this 'peculiarity' of biological free energy conversion, and what it implies for the feasibility of life altogether. But it comes down to the fact that, once again, biological free energy conversions are all mediated by macromolecules executing trajectories of macrostate transitions and that these transitions necessarily involve conformational/allosteric changes in the macromolecular engine (albeit changes potentially triggered by, and that in turn trigger, associated chemical state changes). Conformational changes, furthermore, which are necessarily the product of thermal fluctuations requiring a very large number of concerted individual water molecule impacts. In other words, such conformation changes are instances of Brownian motion of protein-sized moieties taking place as "over-damped", biased diffusions in the viscous medium of water as Astumian in particular has emphasized and elucidated (see e.g. [12, 91]). But, as Astumian has also emphasized, this implies that the conformational movements required in life's engines proceed at the so-called 'terminal velocity'; which velocities, most providentially as it develops, are proportional to the thermodynamic force biasing the diffusion. That is, the conversion trajectories of biology are compelled to also be flux-force linear because (at least some of) their individual macrostate transitions necessarily involve terminal velocity rate-limited conformational changes.

Here again, in free energy conversions chemistry must bow to 'machinery'. But now not just as noted earlier in the requirement that a macromolecular engine must be involved to impose the sequence of conditional steps in the conversion cycles called for in free-energy conversion theory. But also, insult to that injury, by having simple chemistry be no longer what determines the engine's macrostate transitions rates and thereby the applicable flux-force relationship.

In a strong sense, then, what has emerged about biological free energy conversions in general is a belated but forceful vindication of Boyer's insight embodied in his 'binding change' mechanism – an insight now seen to underlie all biological free energy conversions. In which as Boyer understood, the 'changes' he invoked were mediated by – and required – orchestrated and contingent sequences of conformational changes in the macromolecule running the conversion. And doing so necessarily, therefore, in a machinery-dependent manner that clearly "transcends chemistry", as Lane expressively put the matter [47]. But this brings us back to our 'chasm' – that between the powers, or lack thereof, of purely chemical processes, and those arguably vastly more Promethean ones made possible by the engagement of macromolecular 'engine' devices; particularly in regards to the powers needed to engender life. A chasm we now see to be in part the same one that so acrimoniously divided Peter Mitchell and Paul Boyer [92]; Mitchell insisting that somehow simple chemical processes (albeit membrane enabled) must be how 'chemiosmosis' drives ATP synthesis; Boyer insisting in opposition, that a macromolecular 'engine', one whose workings depended on the action of moving, shape-shifting parts, had to be involved. And Mitchell was wrong.

We pause now to take stock of where we are, how, and with what provable confidence, we got here, and what ground we have yet to cover.

# 7 A midpoint assessment; the central points in conflict

In the above it has been argued that:

1. Producing particular non-equilibrium, improbable, 'organized' states is the essential challenge that must be met to bring matter to life; and further that generating those particular ordered states is life's central preoccupation – of its metabolism in particular – and the sole purpose to which its free energy input is devoted. In sum, that 'living matter' is, of necessity, in a very far-from-equilibrium, improbably 'ordered', and inherently dynamic state, and is driven to bend all of the 'free energy' – i.e. all of the externally supplied 'order' (Schrödinger's "orderliness") it can access – to that purpose.

2. Organizing matter involves exchanging increments of improbability in one system for those in another; the 'order-from-order' doctrine. And that this specifically entails 'trading' individual thermodynamically-opposed thermal fluctuations, one that decreases the entropy in the system being 'ordered', for one that increases it in another system. Which trading must take place via a cyclic engine-like mechanism that manages the fluctuation trades while also allowing like increments of organization to be accumulated so that a macroscopic increases in a system's organization can be produced. And the need, of course, is not just to lay bricks in their proper place in a static edifice; life needs to precisely assemble, on the fly, great dynamically orchestrated constructions of staggering and precisely detailed complexity; even ones that can fly and steal your lunch.

But, given the radical nature of these views, and the stakes they put at risk, it seems appropriate to pause here and press a bit harder on the main points of divergence between the 'order-based' conceptions put forward here and their counterparts in established, biochemistry-based thinking; the latter then being – on the warrant of the former – either misconceptions or simply fallacies.

# 8 The core misconceptions

Of these, four are worth a second look because of the central role they play in chemistry-based models of abiogenesis – and especially in light of their status there as unquestioned, even unquestionable, truths: i) that metabolism, and life's processes altogether, take place 'near-equilibrium'; ii) relatedly, that in a biological system any needed element of non-equilibrium organization (i.e. the product of an endergonic process) arises through catalyzed mass action chemistry via a 'coupling' reaction enabled by the transfer of energy between two processes; in which, moreover, the joint process is only mildly exergonic and is therefore just a quotidian example of a near-equilibrium metabolic process; iii) that chemical systems, properly constituted, catalyzed, manipulated, and fed "energy-rich" compounds, or 'free energy' in some other form, can self-assemble, even self-purify and self-concentrate; iv) that to launch, life needed to import a suite of abiotically-produced biomolecules, e.g. the "building blocks of life" together perhaps with some of these incorporated into linear polymers of suitable length. And that gaining access to these resources, in sufficient concentration and purity, is the key initiating requirement of abiogenesis.

The purpose of this section is to present, for each of these articles of accepted understanding, specific arguments in support of the contention that they are each in their turn fallacious.

## 8.1 That metabolism, and life's processes altogether, take place 'near-equilibrium'

Our effort in this piece is predicated on the claim that non-equilibrium organization is the essential and distinguishing property needed to bring matter to life. More specifically, that 'living matter' is, of necessity, in an extremely far-from-equilibrium, improbably ordered state. A state which, moreover, can only be generated by macromolecular 'engines' fueled by ordered matter imported from the environment and carrying out order-for-order, thermal-fluctuation-level trades. Of course, these and the rest of our claims, along with their embedding 'order' verbiage, are irrelevant smoke and mirrors if living systems are really the near-equilibrium, 'just catalyzed chemistry' processes they are conventionally believed to be. So for the present piece, this is a live-or-die question.

What arguments can be marshalled that show with sufficient force that the 'near equilibrium' idea is fallacious? In other words, is living matter provably in an extremely far from equilibrium state?

Two approaches to this question are offered next. In the first we consider what can be said quantitatively about how ordered, how far from equilibrium, actual biomass is. This discussion rests on the fact that we know quite closely how much order is being dissipated during either the creation, or the maintenance of, for example, an E. coli cell, and that this places an upper limit on how ordered the matter comprising that cell can be. We then adduce evidence that on a mass-specific basis, the actual order of living biomass is a substantial, and estimable, fraction of this ceiling value. In the second we present evidence that furthermore, this realized biomass-specific degree of organization appears to represent not just a property of a particular cell type, or even of a particular form of life, but to a very surprising approximation a universal quantitative property of
all extant living systems – that is, of all metabolizing biomass. This implying that to a
stunning approximation, not only is all metabolizing biomass organized to the same
degree but that this in turn suggests that a unit of biomass can do no better.

### 8.1.1 How ordered, how far from equilibrium, is living matter?

Consider again the humble *E. coli* cell. As we have already noted, in a well aerated
minimal salts medium augmented with glucose, it will give rise to a daughter cell in
about 110 minutes. There are certainly not many from the fields of professional biology
who would contend that the atoms in that new cell, almost all of which were minutes
earlier in just a random solution of small molecules, do not now find themselves in a
vastly more organized, vastly more improbable state – now re-sorted and re-arranged
into an immense, very specific inventory of new molecules, great and small, themselves
organized, both structurally and behaviorally, to a degree to shame mankind's best, and
whirring away in the frantic, intricate, precisely orchestrated and purposed endergonic
exertions of biological work.

So, just qualitatively, that Coli cell looks to be pretty extremely organized, pretty
extremely far from equilibrium – and therefore 'way the hell out' on Boltzmann's limb
of improbability. And what is true for that cell, seems certain to be true of them all.
Given Boltzmann's illumination of what order is, what it costs, and in what currency
the cost must be paid (cf. The Introduction), we know that the 'immense' reduction in
probability involved in moving the atoms in the medium to their positions and roles in
the cell, must have exacted a price in the form of an even greater (factor) increase in the
probability of some other matter. A price paid, moreover, by the real-time 'trade' of one
thermal fluctuation creating order, for one destroying it.

It is staggering to contemplate that it is through such 'accidental' and miniscule
dobber-wasp increments of ordering matter that the astronomically far-from-equilibrium
cathedrals of life are created – at the expense moreover of even greater ones dismantled.
But still, in quantitative terms, how ordered, how improbable, how far from equilibrium,
is the new cell's organization? Moreover, is that extent of organization in any sense an
upper limit – perhaps one that is a universal property of living matter?

As was just noted, we can at least put an upper limit on the extent of the
organization of the matter in the coli cell because on the one hand the improbability
gained in building the new cell cannot be greater than the improbability sacrificed,
through order-for-order conversions, to build it; and on the other, we can quite closely
estimate the magnitude of that sacrificed improbability. In, for example, our coli
replication model, there is only one source of the order whose sacrifice can be put to
that service. Namely the disequilibrium embodied in the glucose v. oxygen redox couple
– and which is available to be controllably dissipated, as needed, in the redox transfer of
electrons from glucose to oxygen. In this downhill transfer the state probability of each
electron increases by a factor of $\approx 10^{19}$ (corresponding to a free energy drop of
$\Delta G = 43.5 \, kT$) [93]. And since it is experimentally observed that during replication the
coli cell dissipates the disequilibrium of about $10^6$ electrons/sec [94–96], it will, during
the replication, have done the same to roughly $10^{10}$ electrons, for a net gain in their
state probability by a factor of $\approx (10^{19})^{10^{10}} = 10^{19 \times 10^{10}} = 10^{1.9 \times 10^{11}} \approx 10^{10^{11}}$. As has
been noted, this factor must be an upper limit on that by which the order (the state
improbability) of the matter comprising the new cell has been increased – during, that
is, the replication process. Although, of course, this is a measure of the order dissipated
through conversion to order in the work of assembling the new cell, and much of that
should be assumed to be used only transiently (in maintenance but also construction
costs *per se*, i.e. in importing and processing 'construction' material, processing and
exporting waste), so that by no means does all of the improbability used end up in the organizational state of the new cell. However, that much of it does is suggested by the fact that the same cell building feat, if carried out in enriched media (e.g. "Luria broth"), takes a mere 20 minutes, i.e. is more than 5 times faster if it is spared much of the effort needed to organize the coli cell starting from minimal media [14].

Admittedly, these magnitudes stretch credulity, both as to their literal validity and their interpretation. Is there any literal sense to these huge numbers – and the interpretation being given them?

One way into this question comes from considering the fate of individual electrons in the $\approx 10^6/sec$ flow powering the replicating coli cell. First, by noting that this top-level source of order for the cell is dissipated, in being converted to subordinate disequilibria, as each electron cascades from glucose (mostly via NADH) to oxygen in the cell's electron transport chain – each, as was just noted, experiencing therein a gain in its state probability by a factor of about $10^{19}$. This passage of electrons from glucose to oxygen is the total input of order into the cell; from which fuel, via order-for-order conversions, all of the cell's internal endergonic (order-producing) processes are driven. Second, that almost all of these downstream 'work-a-day' order conversions take place through the intermediate ordered state of an [ATP]/([ADP][Pi]) disequilibrium – typically maintained by cells at about $10^{10}$ (i.e. at a 'free energy' $\Delta G \approx 23\,kT$). This conversion of 'ordered electrons' to 'ordered ATPs' takes place on a just-in-time basis (the average time between ATP synthesis and use being one to two minutes [97], and with a conversion ratio of roughly one electron for one ATP [98]. And finally that the conversion of the ATP disequilibrium to the order-producing processes of the cell, also takes place one ATP at a time (usually that is; 'double barreled' ATPases are invoked for really hard jobs – as we will see illustrated below in the example of nitrogenase); and which exploits not the 'energy' released in the hydrolysis as we will see in detail below, but only the relative irreversibility of that hydrolysis to provide a 'latch' which serves to trap the desired order-producing fluctuation.

Over all, all of the incoming order the cell manages to 'suck in' from the environment is used, under precise fluctuation-by-fluctuation control, in conversion processes to drive endergonic processes and create and maintain thereby the required ordered states in the cell; albeit order, of greater or lesser transiency and created and used in real time: order-dissipated in dynamic balance with order created, electron-by-electron. Thus in these individual electron terms it is clear that if the increment of order produced by each electron powering the endergonic processes of the coli cell is sufficiently persistent and introduced just when needed in the work of assembling the cell (and if both the conversion 'losses', and the temporary/transient expenses of the ordering process itself, are not too great; see discussion of these points below) then the Boltzmannian improbability produced during the construction of a new coli cell may plausibly approach the theoretical upper limit given above. Which value might then be a defensible order-of-magnitude measure of how ordered/improbable the organized state of the coli cell's matter is. A degree of organization which, in light of its cost, might prove to be not greater, not further from equilibrium, than it has to be to bring that matter to life and hold it there (a possibility touched on by the next point).

In any case, hearkening back to Boltzmann's relation (Eqn. 1), each electron delivers, as has been noted, a probability gain ratio to the cell's matter of about $10^{19}$ (equivalently, a free energy loss of $\Delta G/k_B T \approx 43.5$) [93]. And out of that improbability flux all of the raising of the cell's matter 'from the dead', by suitably organizing it, is driven.

---

[14] A stunning fact given that it takes the coli cell twice that long just to replicate its DNA – which fact in turn implies that a coli cell goes to extreme lengths to replicate itself in the shortest possible time – and that, as we might all expect, 'life is hard'.

### 8.1.2 Life's organization ceiling

A second argument favoring the idea that the estimate for the upper bound on the (Boltzmann) organizational improbability given above for a coli cell should be taken literally, is that it apparently reflects, on a biomass-specific basis, a universal limit on how organized living matter of any type can be. To see this consider again that for all of life the overall conversion of the order imported into a cell into that of the cell itself passes almost quantitatively through the same intermediate 'ordered state' noted above. Namely that of ATP driven far-out-of-equilibrium ( [99]) through its synthesis, and near immediate use, against a 'head' of $\approx 10^{10}$. The maintenance of an E. coli cell entails a rate of ATP turnover that is roughly 20-fold less than it is under cell replication [96]; implying a turnover of about $\approx 5x10^4 ATPs/sec$ just to keep an $E$ coli cell afloat. From the perspective of living systems in general this proves to be quite a revealing number. To see why, consider next *homo sapiens*. It is known that just to stay alive we "great apes" turn over roughly a body weight of ATP daily [100]. For the average $\approx 70$ Kg human this is 120 moles/day = $7.2x10^{25}$ molecules/day or $10^{21}$ molec/sec; namely $2x10^{16}$ times larger than the ATP turnover coli requires for the same maintenance function.

Arrestingly, this ratio is also 'exactly' the ratio of their biomasses! So, quite incredibly, on a biomass-specific basis the 'out of equilibrium' maintenance costs for a human and an *E. coli* cell, are the same! This bizarre coincidence must, and in fact does, reflect a much deeper, albeit entirely unexpected, property of nature. And indeed, as Beratan has remarked [94] in summarizing the key finding of the important work of Makarieva's group [95, 96],the biomass-specific rate of electron flux through a cell's electron transport chain needed just to maintain the biomass is to a rough but profoundly meaningful approximation a universal (as must therefore also be its ATP turnover rate). A universal that quite incredibly spans 17 logs of organism biomass, all 'energy' metabolisms, all life circumstances, strategies, and designs – whether prokaryote or eukaryote, single-celled or metazoan, motile or sessile, parasitic or autonomous, plant animal or fungal. Everything living [15].

Why such a universal limit exists, much less it's magnitude, remains unexplained. But the existence of a universal biomass-specific ATP turnover rate strongly implies the corresponding existence of a universal ceiling on how far from equilibrium biomass can go in organizing itself; or, referring back to the earlier quote by Schrödinger, a universal limit on the rate at which a unit of active biomass can "suck" in "orderliness" from its environment and convert it, via order-for-order exchanges, to the ordering processes needed to maintain its own internal ordered state. And keeping in mind that every ATP produced is consumed in an order-for-order exchange driving an entropy-decreasing, order-producing, endergonic reaction [16] it seems that biomass *per se*, from any source, is simply unable to order itself (on a biomass-specific basis) to a degree greater than that universal limit – which limit , moreover, it always, again roughly speaking, attains.

But just noting this last point does it no fit justice. Clearly, the inference that living systems always do drive themselves up against that improbability ceiling fairly shouts that for all of life, being as organized as possible is the key and controlling secret of success. The details of how it is done and how it is used being 'only' context-dependent means to the end. We are then left with the inference that *to the better organized go the*

---

[15] We note that this conclusion, which we here take to be correct, conflicts with other published work; for example that of Delong et al: DeLong J, Okie JG, Moses M E, Sibly RM, & Brown, J. H. (2010). *Shifts in metabolic scaling, production, and efficiency across major evolutionary transitions of life*. Proceedings of the National Academy of Sciences, 107(29), 12941-12945

[16] In the hydrolysis of ATP in endergonic kinase reactions (e.g. hexokinase), the fact that the phosphate is transferred to the other reactant rather than being released into the cytoplasm, only modestly reduces the order dissipated in the reaction – the difference contributing, nonetheless, indirectly and downstream, to the overall order creation work of the cell.

*spoils – if, that is, 'better' means a greater ability to "suck in" and convert order from the environment.* And that life is, at root, a struggle to out-organize the competition – the better to exploit the accessed environment. But a struggle wherein increased organization inevitably confronts a limit at which the next increment fails to add to the profits of organization more than it costs.

As to how such a cost/benefit tipping point might arise, it seems possible that it reflects the intersection of two 'facts of life'. First, it obviously takes organized matter to organize matter; indeed a substantial fraction of the ordered aspects (structural and otherwise) of living matter is clearly taken up, either directly or indirectly, in the order-for-order conversion processes needed to build, maintain, and drive the machines carrying on these very processes – along with, of course, all of rest of the activities and infrastructural elements of the cell needed to support the ordered edifice of the living state. And it seems intuitively plausible that the organization required to generate and support itself, does not scale linearly (on a per unit metabolizing biomass basis).

Second, the inherent instability of highly organized states, especially those with substantial interdependence of their components, increases non-linearly with the system's complexity. For these reasons the complexity of highly interdependent organized, self-assembling systems, such as cells, cannot increase indefinitely – and must presumably reach a limit at which the next incremental increase in complexity only just balances, on average, the added order-consuming cost of 'keeping the whole operation together'. And as we noted above, the data of Makarieva et al [96] imply that the ATP turnover cost of maintaining biomass is about 5-10% of the cost of assembling it in the first place. This in turn suggests that the rate at which the cell's supply of order must be invested in just repair and homeostasis, implies that, absent that investment, cells typically 'go flat' and fall apart, with a half-life that is only a couple of logs greater than the time it took to build the cell in the first place.

In summary, the above conclusions are that a ceiling can be legitimately calculated on how organized, in Boltzmann's terms, biomass can become; that the level of organization attained by metabolizing biomass is to a stunning approximation a 'universal' (biomass-specific) number; and that phenomenologically this universal level stands not too far below its theoretical upper limit value. We take it that together these considerations support the claim that the processes of life, those of its 'metabolic chart' included, are certainly not in any valid sense operating in equilibrium's shadow, but instead quite extremely far above it. And do so, it seems hard to deny, because the biomass could not be alive otherwise. This also making it obvious that the widely held consensus to the contrary rests largely on the root misconception that the coupling of exergonic and endergonic processes is an instance of a (suitably catalyzed), mass-action chemical reaction – operating, moreover, by the transfer of energy between the two.

In this misconception it is this energy transfer idea that most fatally, and insidiously, offends the truth. This in part because it seems an obvious certainty to all minds lay and learned – describing all processes doing 'work' of any kind. And in part because it is cemented into the bed rock of unquestioned verities that form the conceptual cannon of the entire dominion of the life sciences, biochemistry most centrally, and spanning the entire history and reach of the subject. For that reason we discuss the energy transfer idea at greater length directly below (Sec. 8.2). But first address a point related to the 'near-equilibrium' idea: namely the fact that although coupled reactions (free-energy conversions) do in general run close to the state in which the entropy being created in the driving process is not too far above the entropy being lost in the driven one, there is very important 'biology' in exactly how far the former exceeds the latter.

### 8.1.3 In a free energy conversion, by how much should the exergonic process's entropy production 'over-power' its endergonic partner's entropy consumption?

The observation that in coupled reactions the net change in free energy is small and that therefore they can be fairly regarded a 'close to equilibrium' processes is at best misleading and at worst misses an important point as to how bioenergetics works; namely the question of what determines just how far from being an equilibrium process 'should' free energy conversions be made to operate? It is trivial to note, of course, that coupling processes cannot be running too close to equilibrium; when the two processes are in thermodynamic balance no net conversion, no net work, is done. But there is more to it than that. In fact, the degree of imbalance in a conversion process is of substantial quantitative importance itself because it effects two critical performance measures of the conversion: 1) the speed with which the conversion process operates; relatedly, and likely more importantly, the power output of the engine; and, somewhat separately, 2) the degree to which the conversion engine's efficiency is compromised by waste cycles. This presents an interesting pair of questions: first, is there an 'optimum' value for the ratio of the rate at which the driving process is producing entropy to that at which the driven one is reducing it? Perhaps, for example, a value which maximizes the latter rate, i.e. the rate at at which the engine is doing the work which is, after all, its *raison d'êtra*? And second is this optimum ratio actually found to describe the operation of biologically important conversions?

The general answer to the maximum power question, under standard simplifying assumptions, and for conversions involving processes in which the fluxes are proportional to the forces – as is the case for all of those of life [62] (and, interestingly for classical heat engines as well) is remarkably simple both conceptually and in formal terms; namely that: *the power output of the engine is maximal if the rate of entropy production by the driving process is, in magnitude, twice the rate of entropy reduction in the driven one.* See Appendix C.1 for arguments supporting this assertion.

But does this simple rule fit biological reality at all? Consider the case of the 'master conversion' of life, that turning the redox flux to which, universally, order imported from the environment is initially converted [99], to the order of life's primary and internal chemical fuel, namely the $ATP-v-ADP$ disequilibrium – of, that is, the process conventionally called oxidative phosphorylation (Ox-Phos). And for that case we happen to have the answer directly at hand. Specifically, we had earlier used the literature values of $-43.5\,kT$ and $+23\,kT$ for the free energy changes experienced, respectively, by an electron descending the ETC chain in mitochondria, and the synthesis of an ATP by ATPsynthase – a conversion which we noted takes place on an approximately one electron for one ATP basis. And though we are not quite at entropy changing rates yet, recall that for processes in which Gibbs free energy is well defined that $\Delta G_{system} = -T\Delta S_{system+bath}$ [3, Ch. 8]. Therefore, in Ox-Phos, life's top level order-for-order conversion, the ratio of entropy gained in the driving process to the magnitude of the entropy lost in the driven one, is here estimated to be $43.5/23 = 1.9$ (!). Considering the approximation assumptions and somewhat arbitrary choices of reference values involved, that this number is so close to the theoretically magical 2.0, must be fortuitous. It nevertheless seems to suggest with some force that life has seen to it that Ox-Phos is running close to its maximum power output; at that level of coupling where the rate at which the conversion is producing order embodied in an ATP disequilibrium is maximum. Which 'order', of course, is quickly, one could fairly say ravenously, consumed in down-stream conversions driving the creation of the subordinate non-equilibrium states which in their aggregate bring the matter being organized to life.

All this offering further evidence, we suggest, that quasi-equilibrium, mass action chemistry, however catalyzed, is a model of life's molecular processes which stands as far from its reality as the organization of a coli cell does from that of the solution of small molecules from which it was assembled.

## 8.2 Coupling by energy transfer: That the coupling of exergonic and endergonic processes is an instance of a catalyzed chemical reaction which operates by transferring energy between the processes

As was noted above, particularly among biochemists, the foundational assumption in the accepted understanding of how the 'coupling' of an endergonic to an exergonic process operates to achieve free energy conversion is that energy released in the exergonic process (e.g. the hydrolysis of ATP) is transferred to, and drives, the endergonic one (by then rendering it in effect exergonic). This, as we have noted multiple times is not correct; but from this error a second follows. Namely the idea that via the imagined transfer of energy the two processes are thermodynamically fused into a single chemical reaction, one which can be understood in standard mass-action, Michaelis-Menten, 'just catalyzed chemistry' terms. This notion supports in turn the more general inference, also incorrect, that the transformations of metabolism, its free energy conversions most centrally, comprise a network of chemical reactions all operating 'near equilibrium' – and needing, moreover, no more essential macromolecular help than enzymatic 'catalysis'; see e.g. [101].

In the preceding sections we have argued instead that: free energy conversions do not operate by, or involve, the transfer of energy between the partners in the conversion, are not 'just chemistry' no matter how catalytically assisted they are imagined to be; and are neither operating 'near equilibrium' nor simply instances of 'mass-action', chemical reactions [17]. But that instead, and in principle (see Sec. 6.1)), endergonic, thermodynamically uphill processes can only be made to happen by 'trading' (and accumulating) individual, thermal (Brownian) 'uphill' fluctuation events in one system, one at a time, and *after* they happen by chance – for individual fluctuation 'downhill' events in another system. To make this trading happen the processes involved must be under the control of particular type of macromolecular device managing a number of distinct macrostates of the system together with stochastic trajectories of transitions between them – as described in Section 6.1, and in Table 1.

In this section we draw attention to the above assertion that in the fluctuation-trading process, the exergonic (driving) fluctuation typically occurs *after* the endergonic, order-producing fluctuation, itself the result of a chance Brownian event, has taken place.

How could this 'pay after completion' ordering of events possibly be true, since it seems to flatly contradict the 'God given' 'energy transfer' concept? However, notwithstanding this affront, this seemingly paradoxical driver-after-driven ordering of the events in a free energy conversion cycle is neither new nor, among experts, controversial; having been advanced, as we have also noted, in the 1990's by the eminent biochemist William P. Jencks [82]. Moreover, it is a prediction confirmed, to my knowledge, in all cases where the order of the events has been determined (see discussion in [90], and the nitrogenase example discussed next). Furthermore, again to my knowledge, in all cases where the mechanism involved has been sufficiently explicated, this order of events is achieved by having the occurrence of the endergonic event 'trigger' or 'gate', the exergonic one.

---

[17] No less authorities on the matter than Hill and Eisenberg wrote a short note in 1981 aimed, and with no little asperity, at purging the energy jumping idea; arguing instead, and correctly, that it is the transition probabilities between the stations in the entire conversion cycle that effects, for purely stochastic reasons, the conversion [141].

Together, these observations suggest first that, as we have argued, the energy transfer idea is wrong, and second that what the exergonic event does do, namely prevent the reversal of the endergonic fluctuation event, is much better achieved if it is mechanistically gated by, and thus occurs after, the endergonic event; the system then 'knowing' that there is now an unstable but valuable new state that needs quick saving. And in fact, not only does energy not change hands in the coupling process, neither is it in any sense 'conserved'. Instead, the more fully the *free energy* in the exergonic process is dissipated, the more irreversible it is, and thereby better able to perform its jobs of: functioning kinetically as a latch to trap an uphill fluctuation, and 'thermodynamically' to balance the 2nd law books.

It might seem that observing this 'paradoxical' ordering of events, in which, for example, the driving hydrolysis of an ATP molecule is made to occur after the work it is 'driving' has taken place, would have quickly put the end to the energy transfer idea. But that is not at all what has happened. Instead, impressively, experimentalists stumbling onto this driver-after-driven sequencing fact have, to my knowledge, shown no sign of considering the possibility that the observation vitiates the 'energy transfer' idea. But have instead nimbly proposed that there must be some short-term borrowing mechanism in play; which they have variously termed "credit-card– or "deficit financing– bioenergetics"; no further questions being, or need be, asked . The existence of a deficit financing mechanism being now taken as an established concept (again, 'what else could it be?'), often just noted in passing; in spite of there being no evidence for it, much less any mechanism proposed to support it (see discussion in [90]). And of course, there is no such thing; nor is there the slightest need for it; because the energy transfer idea is just flatly wrong.

But of course, it's not just the energy transfer notion that falls here. So too its companion: that the free-energy-converting couplings of biology are just catalyzed chemical reactions. To the contrary, as this piece has sought almost single-mindedly to establish, they are not; instead, an engine is required, and for inescapable thermodynamic reasons. Indeed, in every sufficiently characterized biological free energy conversion, a macromolecular complex functioning as an engine has been shown to be involved. [18]. A stochastic engine, moreover, managing individual thermal fluctuations in the reactions being coupled; such that key fluctuations in the two processes are made to march in an alternating thermodynamically opposed manner; the occurrence of a desired endergonic fluctuation in one of the two processes being then used by the engine to gate, i.e. not to cause but to permit, a matched exergonic fluctuation in the other – while at the same time striving to ignore and/or block very nearly all of the other thermal fluctuations in the system that Dr. Brown puts on offer. All of this to-ing and frow-ing just to convert the dissipation of order in one system into the creation of order in another – where most often, notably, the two systems are completely unrelated in chemical terms. And in the absence of the engine, would each proceed exergonically, independently of, and unaffected by, the other.

In this, as we have noted above, the engine's job is to see to it (ideally) that the 'neither-unless-both' rule is enforced in each cycle: i.e., that a given engine cycle should be allowed to proceed to completion if and only if in that cycle one fluctuation-mediated instance of the process being ordered is paired with one of the process being disordered. For good engineering reasons (not ones of thermodynamic principle), this requirement appears to be more-or-less invariably achieved by the engine using the occurrence of the

---

[18] Just, it is worth recalling, as it was found by Paul Boyer to be true for the 'index case' of ATPsynthase. This 'queen-mother' free-energy converter, he showed, operates by a conformation-shifting 'binding change' macromolecular, cyclically-operating mechanism; one acting as a true engine in the sense intended here. A profound Nobel-garnering insight; notwithstanding that it was furiously opposed by Peter Mitchell; who, as we have noted, insisted in opposition, and till his death, that pure (catalyzed) chemistry had to be, somehow, sufficient [92].

endergonic fluctuation to trigger its exergonic mate – which must then take place
after-the-endergonic-fact, and as quickly as possible. Thereby allowing a sufficiently
irreversible exergonic event to best function as a reliable latch to prevent the loss,
through reverting, of a single lucky, anti-entropic fluctuation.

To make this heterodox view of things more concrete a summary description of a
particularly illustrative, and amazing, example is discussed next: that of nitrogenase;
the machine that manages the 8-electron reduction of $N_2$ to $NH_4$ (an example whose
singular merits we have also noted in earlier publications; see e.g. [62]).

### 8.2.1 Nitrogenase

We here rely on the remarkable 2013 paper by Duval et al. whose key result is given the
emphasis it deserves in the paper's title: "*Electron transfer precedes ATP hydrolysis
during nitrogenase catalysis*" [102].

Attention is drawn in particular to Figures 1 and 4 in the paper. In simplified brief,
the system executes an elaborately choreographed cyclic conformational dance involving
two proteins: the "FeMo" protein whose key elements are an FeMo cofactor which binds
$N_2$ and the partially reduced intermediates in the sequence $N_2$ to $NH_4$, and mediates
the one-electron reduction transitions in that sequence – together with a nearby [8Fe-7S]
one-electron redox center; and the "Fe" protein which is a homodimer ATPase bearing
two nucleotide active sites and a single [4Fe-4S] redox center. Each one-electron cycle
starts with the FeMo's redox center reduced and the Fe protein in its fully loaded state
(two ATP's bound and its redox center reduced) docked to the FeMo protein. In that
state the essential and very strongly endergonic fluctuation producing the transfer of
one electron from FeMo's redox center to the FeMo cofactor can, by Dr. Brown's
agency, occur. When it does, a fast but relatively weak 'backstop' reaction takes place
in which the FeMo's just-oxidized redox center is reduced using an electron obtained in
a redox transition from the Fe protein's redox center – thereby blocking, at least
transiently, the reversal of the one-electron reduction of the cofactor site. This state
then triggers a slower but much more powerful backstop reaction in the form of the
'double-barrelled' quasi-simultaneous hydrolysis of both ATPs in the Fe protein. At
which point the now fully discharged Fe protein dissociates from the FeMo protein (to
be recharged elsewhere) – and the single-electron cycle can be re-initiated. The evidence
presented in this remarkable paper renders its title assertion incontestable: that in this
very dramatic example, the general, orthodoxy-confounding fact is that in each cycle of
the coupling reaction the endergonic work is done before (and triggers) the exergonic
reactions that trap and pay for it.

As we have noted, this 'inexplicable' ordering of events has been observed in a
variety of different systems, in fact, to my knowledge in every case where the question
has been addressed (a list of examples is given in [90, Sec. 11]). But never, to my
knowledge, have the investigators observing it suggested that this throws the least shade
of doubt onto the 'energy transfer' theory of coupled reactions. Instead, as we noted
just above, it has been uniformly proposed – in some cases even taken to be established
knowledge – that some kind of short-term energy lending mechanism must be involved.
And proposed with no evident discomfort with the idea, nor any expressed concern to
fathom in mechanistic terms how biology could be, in just these conversion situations,
'both lender and borrower' (notwithstanding the biblical injunction to be neither).

But of course, as we have also noted, the pursuit of this presumed
delayed-gratification energy is chasing a phantom. There is none to be found, nor any
needed. To repeat for emphasis, the energy transfer idea is just false.

## 8.3 Self-assembly and self-concentratiom/purification

As we have also previously noted, the idea that life needs to have abiotic processes of nature supply a certain kit of 'bio-molecules', or at least of 'bio-like-molecules', is an inescapable tenant of 'chemistry-based' OoL thinking. And all adherents thereto are excited by any finding of some of these desired molecules in the products of any synthetic process, whether of the lab or nature, of terrestrial or astrophysical provenance. Excitement impressively undiminished by the fact that in all cases, the desired biomolecules are vanishingly small 'contaminants' of the products produced, or present at vanishingly low concentrations, or both. This is a strange fact indeed since no biochemist would try to make any sort of specific chemistry commence starting with reagents and reaction conditions anything like these. Implying, of course, and with no escape, that somehow corresponding processes of selective purification and concentration must be brought to bear; necessarily ones of extreme selectivity and specificity. All such tasks requiring that an improbable, 'ordered' subset or version of the original state be produced. To which Boltzmann demurs: a) 'show me the ordered material, whose disordering, one thermal fluctuation at a time, can be made to be specifically conditional on, and possibly thereby 'trap', individual chance endergonic fluctuations in the given sample which move it an increment, specifically towards the state desired; and b) show me the the device that will actually enforce these conditional fluctuation trades, and also make it possible for favorable increments of order to be individually accumulated. Of course, 'Boltzmann's' questions are facetious, since he knows that in principle, 'chemistry' (absent any of the needed, highly specific conversion engines) cannot provide an answer.

One imagined escape from this dysphoric view still has general currency however, particularly among RNA enthusiasts: the imagined ability of autocatalytic processes to arise spontaneously and impose order that was not there before. But this idea too is inherently just magical thinking. The order-only-from-order reality guarantees it. First, just as before, what ordered system is it whose order can be dissipated, in thermal fluctuation increments, each one contingent on the immediately prior chance occurrence of an order-creating, anti-entropic fluctuation in a particular RNA oligomer being autocatalytically self-organized? Second, what process or device will impose the contingent dependence of the needed exergonic fluctuations on the endergonic ones they are each being traded for? Do 'self-organizing-dissipative structures' ("SODS") offer an out? That is, assuming the system in question could be specifically placed in a strongly non-equilibrium state with respect to one of its 'chemical' degrees of freedom, could a SODS arise spontaneously whose formation facilitated, and itself embodied an organized structure of biological utility? In the section below "Why can't life use SODS internally ... ?" (Sec. 10), we attempt to justify an unequivocal negative answer to this question.

That is, we claim that chemical processes on their own just cannot organize matter in the ways and contexts needed to facilitate abiogenesis. And even if they could, the system would then surely move, would "devolve" away from rather than evolve toward, useful organized states; moving ineluctably into the deadfall trap of "Benner's curse"; whereby the endpoint of any and all sequences of chemical 'free energy excited' manipulations is entombment in the frozen disorder of asphalt.

## 8.4 The necessity of importing 'life's building blocks' and perhaps 'polymers' thereof.

Biochemists interested in the EoL problem are more or less uniformly persuaded that key to igniting abiogenesis is having a solution of either the right mix of 'life's building block' molecules, or credible precursors to them, and also, many believe, suitably long polymers thereof. The hunt then being on for an abiotic process that can produce a plausible starter kit of building blocks (BB) in useful purity, concentration, specificity and (if needed), 'activated' state – and/or for one in which, given a supply of polymerizable BBs, and feasible abiotic manipulations, interestingly long linear polymers thereof would form spontaneously. It being axiomatic that then biopoesis ignites and 'life' emerges autocatalytically [19] [103–108].

That is to say, given a process that can produce a solution of such 'building blocks', or polymers thereof, augmented as one pleases by catalysts, reagents, and perhaps special surfaces, supplied with 'energy' in some form, and perhaps 'cycled' in some manner (e.g. in water activity or temperature), then, it is thought, life will self-organize and rise up.

However, here the same vitiating considerations as we have just invoked regarding purification and concentration also apply; independent of whatever set of blocks, or polymers, one specifies: matter can just not be made more organized in such ways; Boltzmann's order-disorder principle forbids it. To perhaps see this a little more concretely, consider the 'polymers first' idea. In the OoL context only heteropolymers are of biologically-specific interest; and ultimately of course, biology is only interested in heteropolymers whose monomeric sequence is more or less strictly determined. That is, biology is only interested in a small (in practice astronomically small) subset of heteropolymeric sequences of a given length (and with equal passion for not having all of the other sequences lying around). But, of course, chemistry can only generate pseudo-random mixes of heteropolymers. Throwing us back, at a minimum, either into the purification/concentration deadend trap we discussed above, or to the false belief that one particular sequence might arise that could self-amplify and, gobbling up all the monomers, somehow initiate the abiogenesis of biological organization. But this again is a notion Boltzmann does not permit us to entertain because of the fundamental mechanistic and thermodynamic requirements discussed earlier regarding how improbable, ordered, outcomes are generated – such as sequence-specific linear heteropolymers.

Of course, the 'too many different sequences' problem that arises in the chemical syntheses of heteropolymers, is just an example of the broader, "combinatorial problem" which advocates of the chemical OoL models have long realized they face. We briefly take that up on its own terms next.

### 8.4.1 The 'combinatorial' problem through Boltzmann's eyes

From the beginning of serious work on biochemical approaches to abiogenesis, stimulated by the Urey-Miller experiments [109], it has been recognized that such approaches are plagued by the so-called combinatorial problem; i.e. they unavoidably produce too many, generally vastly too many, molecular species. And while molecules life uses may be among them, they are never more than minor contaminants. And on top of that, the approach unavoidably faces a Benner 'tar-pits' problem: the more free energy is supplied, or more chemistry methods are invoked, the worse it gets – as is called out, and given a literally attractive asphaltic twist – in the BKC paper discussed above. In summary, the 'combinatorial problem' starts out unacceptably severe and, on the warrant of BKC, can only get worse if further free-energy supplied procedures are pursued.

However, in significant contradiction to the BKC assertion, much effort has been and continues to be expended seeking chemical or physical-chemical ways of reducing the number of molecular species present after some chemistry has had its synthetic way.

---

[19]The author hopes he may be excused for not affecting to overlook the glaring 'cargo cult' character of this view of things: "*If we build us a runway and put lights on it and ....*" – where, of course, what matters is in no way just the obvious 'otherwordly' structures, such as linear polymers – much less sequence-specific heteropolymers – involved in the biological deliverance of cargo.

Efforts, moreover, which have reported some (modest) success [46, 104, 110]; on the basis of which it is generally proposed that, sufficiently repeated, such methods would yield species mixes that were not just simple enough, but ones specifically enriched for just those species from which life, it is presumed, could – and would – then 'self-organize'; [20] (this incidentally raising the question – not further addressed here – of whether the BKC 'asphaltic death' assertion is wrong, or the methods used to estimate molecular species complexity in these hopeful experiments are not measuring it in a thermodynamically relevant sense). In any case, if 'chemistry' EoL models are to be taken seriously, the combinatorial problem must be solved. Something must stand ready and able to purify out of the extremely diverse mix of molecular species chemical protocols inescapably produce, an extremely specific, very minor, and improbable, subset of that diversity. Based on just the methods of chemistry, is this even possible in principle, much less something that could happen naturally? In other words, can chemistry rescue itself from its own combiniatorial problem? As has already been discussed (see Sec. 8.3), it cannot. And for thermodynamic, not practical, reasons.

That is, chemistry processes aimed at producing credible starting conditions for life's emergence, inherently confront a 'fatal' purification/concentration problem, one which chemistry methods cannot, in principle, solve. Of course, such purification and concentration tasks can be carried out by going 'full Boyer': i.e. wheeling up true macromolecular order-converting engines, plus the specifically ordered matter needed to fuel them, and membranous structures to distinguish 'in' from 'out'. But 'Boyer' involves highly ordered, improbable configurations of matter, including order converting engines, and clearly cannot arise by the arts of chemistry alone. In fact, we claim, life certainly did not start this way; did not start from a terrible chemical mess which then magically purified and concentrated itself. The 'self-purifying mess' concept is not, quite literally, 'viable'. But then what else is there? What other, fundamentally different approach exists that avoids these (and other) fatal problems inherent in 'chemistry' approaches? There is one, one born of naturally-arising abiotic engines that can organize matter, and in just the right way from life's point of view.

## 8.5 The engines of order: enabling the mess-free (and only) way to create the ordered edifices of life: through a sequence of very small, accidentally produced inventions, each incremental step built on its predecessor, each vetted for fit and suitability specifically in the context in which it arose.

The only path to glory here is a mess-free one; starting simple and clean, and with the right abiotic engines, fuel sources (e.g. redox and ion gradients), and 'membranes', needed to get an abiotic conversion process going which can generate an initial seed of organization, e.g. formate having been 'fixed' from $CO_2$, and thereafter ascend by incremental elaborations on the theme of this first step to higher levels of 'pro-life' organization; each step 'vetted' in the context of its predecessor. But note that not one even slightly complex invention of mankind has come about in any other way: the cell phone by way of Faraday to Edison to the transistor to the modern miracles of microelectronics, a path had to be traversed comprising innumerably many incremental 'inventions' each tested and tried, by armies of those trying, in its predecessor's context. The invention of life was surely exactly that type of process. And, we conclude this section by noting, that only one of the available EoL concepts is consistent with these requirements; namely the engine-type Alkaline Hydrothermal Vent model advanced by

---

[20] An idea which could nominally be put to a direct test (which it can be guaranteed to fail), by simply buying all of the molecules in purified form from a commercial source, adding any abiotic catalysts one likes, mixing and manipulating them in any way desired, and then waiting.

Michael Russell and his colleagues.

In any case, we are driven again to the conclusion that to both get life started and to operate it ever thereafter order-for-order conversion engines are the essential 'chemistry transcending' ingredients. Thus, we are forced back to the question, which we take up next, of how engines come into existence.

# 9 Where do engines come from?

The engines of relevance to life, to summarize, are macromolecular stochastic devices fueled by the conditional dissipation of order in one system and creating, as the engine's work product, order in another. This in a literal trading of order for order effected at the level of individual thermal fluctuations: whereby a chance anti-entropic, order-creating fluctuation in one system is captured by having its occurrence 'gate' an order-dissipating fluctuation in another. As we have sought to make clear in fair detail (see the Introduction and Sec. 6) this trading necessarily involves a type of precisely choreographed molecular *pas de deux* comprising a disciplined sequence of discrete, fluctuation-level molecular transition 'steps' executing a specific intertwining of the two processes' half-reactions. A dance whose execution requires the mediation of a be-spoke, task-specific, macromolecular engine. And which is by no stretch an instance of mass-action chemistry, however catalyzed.

But suppose it be accepted for the moment that living systems must organize their comprising matter into far-from-equilibrium configurations, and that this takes place by order-for-order conversions that can only be carried out by macromolecular devices operating as engines. Where do the engines – themselves macroscopically organized states of matter – come from (not to mention the externally-supplied 'order' needed to fuel them)? And most particularly, of course, those needed to effect the order-creating transitions required to initiate abiogenesis? This is clearly a key question on which our story hinges.

To open on this point it is necessary to deal first with a fundamental and consequential dichotomy that separates order-converting engines into two distinct classes: spontaneously arising, self-organizing ones, and 'ready-made' ones – a distinction noted in the previously cited paper by Cottrell, "The Natural History of Engines" [15].

## 9.1 Self-made vs ready-made engines

The 'self-made' engines are the so-called "Self-Organizing Dissipative Structures" ("SODS"), first brought to light by Prigogine and his colleagues [111]. These arise in some systems under certain types of thermodynamic stress, forming spontaneously and acting to accelerate the dissipation of the stress producing them. Convective flows (e.g., the earth's mantle plumes ) of all types are paradigmatic examples. But there are many others. Including, importantly, SODS that arise in purely chemical systems; so-called 'reaction-diffusion' structures for example [112–114].

Arrestingly, nearly all of the order-creating engines in the Universe – responsible for virtually all of its wondrously organized structures and processes – are Prigogine 'SODS'. But not all. A tiny fraction of the order conversion business of the Universe is carried out by the ready-made class of engines, as we discuss next. But before we make too much of their relatively negligible quantitative impact we should take note of the fact that all of the engines at work in living matter are ready-made. A point that is trivially satisfied for those of extant life, but presents a decidedly non-trivial question for life at its emergence.

The engines of extant life come into being, of course, via the usual genomically-informed mechanisms underpinning, primarily, protein synthesis. But no less ready-made were, almost certainly, their abiotic forbears – engines which, thermodynamics demands, had to have been operating 'in the beginning' – to organize matter in the particular way needed for it to take its first steps up the flight-from-equilibrium ladder into the organized deliverance of being *alive!* We take up considerations in support of this point below (Secs. 10, 11), including direct experimental evidence establishing its plausibility. In any case, none of extant life's engines are SODS. Nor, we argue below, could they be. Yet, SODS are by no means irrelevant in the engendering and perpetuation of life; indeed, the connection between SODS and the ready-made engines, is a subtle, complex, and interestingly interdependent one. A point we return to below (Sec. 10). And the ready-made engines upon which life relies enjoy another noteworthy distinction. Quantitatively negligible as their impact may be overall, their qualitative significance is another matter. For they, working together in functional collectives, generate what is arguably, and by an immensity of degree, the most extremely organized, the most extremely improbable states of matter (in a mass-specific sense) that exists in the Universe. Namely life. More on this point also below.

## 10 Why cannot life use SODS internally – and of what use are SODS to life?

What exactly is not to like about SODS as mediators of the free energy conversions of life? Let us count the ways:

1. First, none of the central conversions needed by life, especially to launch, are, or can be, carried out by SODS.
   SODS, after all arise out of, and dissipate, their 'own' ordered/disequilibrated state, as dictated by the physics of that disequilibrated system. And produce their own product ordered state, also dictated by the physics and/or physical chemistry of the driving disequilibrium (consider the conversion of a thermal gradient to a vortex convective system). Life, on the other hand, universally converts a limited set of sources of order (primarily redox disequilibria and electrochemical gradients ( [99]), to a limited set of 'ordered', i.e. disequilibrated, states the need for which is dictated by the exigencies of biology; states, moreover, which are generally unrelated chemically to those of the driving disequilibrium.

   To simplify the matter a bit, but without distorting the basic essentials, sitting universally at the top of the order-converting hierarchy of life is first a redox gradient, which is converted to a trans-membrane ion gradient, which is converted to a ATP/ADP concentration gradient, which then is converted (via 'coupling' processes driving various endergonic reactions) to all of the other ordered states life needs to become, and keep itself, 'viably' organized. None of these universal conversions are, nor could be, instances of SODS. And if you are not in those specific conversion businesses, you have nothing to offer life – directly.

2. Second, for the reasons just stated, SODS are self-defining and self-controlling processes.
   Life in contrast needs converters that mechanistically 'separate' the dissipation of the driving order from the creation of the driven one. And in that separation provide the levers of opportunity needed for both design and regulatory control. SODS offer neither.

3. Relatedly, SODS are entirely 'self-serving' engines.
   Being, by the physics forcing them into existence, thermodynamically devoted to creating and sustaining the ordered system that is the SODS engine doing the conversion. That is, the order they generate and sustain is only incidentally anything other than that of the conversion engine creating that order. Life arguably demands the services of more 'selfless' conversions; those whose ordered state output is not primarily (and certainly not directly) that of the conversion engine itself.

4. And then also, there are awkward bits about 'fitting in';
   ... both into the particular chemistry life needs and into, even, the space available. Consider any of the known reaction-diffusion systems; as an essential requirement they work by very specific and complicated reagent-sharing chemical processes; ones which brook no confounding reagents or parasitic paths. In addition, the order they produce is merely oscillating patterns of spatial concentration inhomogeneities. Patterns whose properties have no apparent relevance to life's needs – this only in part because in general their spatial dimensions far exceed those available in the densely crowded precincts of cellular life. Cute perhaps, but no cigar. No such systems are of plausible relevance to life and are never used by it.

However, caution is warranted before accepting this negative conclusion. First, as we have noted, some SODS arise in purely chemical reaction systems, so a pure 'chemistry' model of the EoL might in principle be made to be at least thermodynamically consistent – if it could be shown that self-organizing 'pure chemistry' engines: 1) can arise spontaneously in the life-relevant chemistry under consideration, 2) produce internal order that is not entirely that entailed in organizing the engine itself but that can be in part diverted to other uses without collapsing the SODS, and 3) is of a specific type that natal life could have made use of in order to launch. We claim, however that these conditions cannot be met; the third in particular, whether the first two are or not. And that this is the case not only for the needs of extant life (presenting a possible explanation of why none of extant life's engines are SODS), but for those of its birthing as well.

Yet notwithstanding all this, SODS do play an essential role in life's story, its emergence in particular. As was suggested above, many of the external sources of order (and sinks for 'waste' disorder) upon which life ultimately depends (e.g. those powering autotrophs), and every one on which it directly depended at its emergence, is the providential byproduct of a SODS (as Cottrell discuses [15]). For example, if you fancy (as you should, of course), the alkaline hydrothermal vent ("AHV") EoL model propounded by Russell and co. [115–117], then the driving disequilibria (redox and ion gradients), the structures needed to support both those gradients and also to potentially function as the conversion engines (see Sec. 11 below) are all served up 'right there' in the vent processes, 'for free'; they are all the direct, 'ordered' byproducts of mantle pluming and serpentinization, of, that is, the convective SOD processes of mantle plume tectonics [120] (but see also: [118, 119, 121]); this being arguably a noteworthy, and entirely unique, virtue of the AHV model.

## 10.1 Could inanimate processes have actually provided the ready-made engines, and the fuel they need, to launch life?

If self-organizing free-energy conversions are never used by extant life, and by the arguments just presented almost certainly never were, and if it is accepted that every step in the emergence of life, right from the first, required generating non-equilibrium, organized states, what ready-made abiotic engines could have been available to get life launched? As we have discussed in previous publications [90, 122–126], there is now a very substantial body of evidence that certain minerals, most notably the variable valence, double layer hydroxide green rust, can function as engines to mediate a number of biologically essential conversions [124]. Here, a single example will be reviewed, although it is a stunningly revealing and biologically central one; that of the reduction of nitrate to ammonia by green rust.

## 11 Proof of principle that certain minerals can carry out, entirely abiotically, at least some of the more challenging order-for-order conversions of life.

Arguably, the 8-electron reduction of nitrate to ammonia is one of the most daunting in biology's repertoire. This not least so because of the magnitude of the endergonic lift involved, but also because the reduction must proceed one electron at a time; i.e. as a sequence of 8, one-electron reductions (see [127]). The finding, initially by H.C.B Hansen and his colleagues, that the redox-variable mineral green rust can not only function as the needed conversion engine in this process, but do so rapidly and with almost quantitative efficiency, is therefore nothing short of astonishing (see especially H.C.B. Hansen et al. 2001 [128–130], but also [131]. The reported experiments of the Hansen group are 'one-shot' batch reactions, in which highly reduced GR is immersed in a relatively concentrated solution of nitrate. The oxidation of the GR and the concomitant reduction of nitrate then proceed, and both are assayed during the process. In these experiments, in other words, the GR functions not just as the conversion engine, but also, when reduced, as the source of order whose progressive dissipation thermodynamically drives the endergonic nitrate-to-ammonia process. As this would predict, the velocity of the reduction of nitrate falls as the oxidation of the GR proceeds. At the end of the experiment, the GR is as oxidized as it can become, and the reduction of nitrate has ceased (see in particular Figure 3. in Hansen et al. 2001 [128, Fig 3]).

Arguably, beyond the fact that the 8-electron reduction process takes place at all, the most remarkable and potentially telling thing about it is a further observation reported by Hansen et al. – namely that this conversion process is 'flux-force' linear: that is, the instantaneous velocity of the reduction of nitrate to ammonia is linearly dependent on the thermodynamic force driving it at that moment (which force is itself proportional to the fraction of the redox-labile sites in the GR that are still oxidizable). And this fact is on its own a stunning observation – in light of the point made above that whereas biological free-energy conversions, in stark contravention to standard free energy conversion theory, show flux-force linearity, they do so only because they are mediated by a macromolecule whose operation depends on relative motions of parts of that molecule through the viscous medium of water (see also [62]). This suggests that in the Hansen et al. experiments, the observed conversion of the oxidation of individual Fe(II) sites in the mineral's interlayer galleries to the reduction of nitrate also depends on local conformational fluctuations – most likely acting peristaltically [132, 133] – in the walls of those galleries and therefore necessarily taking place in the viscous medium of water.

We emphasize that this remarkable example establishes two critical 'proof of principle' points. First, that at least one mineral, green rust, can function as a very high quality order-for-order conversion engine; performing, moreover, one of life's most important but also most daunting conversions. Second, that it does so in a manner that is stunningly 'life-like'; including, most unexpectedly, and perhaps most importantly, that of operating in a linear flux-force manner. As we have noted above (and see also [62]), this behavior is not only the most unexpected and anomalous property of life's free-energy conversions, but it is their particular property without which they would not be able to sustain the living state (see discussion in Sec. 6.1).

The ability of Green Rust to be both the engine and the source of 'order' for a variety of biologically relevant reductive conversions is also exemplified by the recently reported conversion of methane into methanol [134], driven by the oxidation of GR to magnetite or lepidocrocite, which ultimately results in the generation of very reducing (strongly out of electrochemical equilibrium) reductants as described previously [135].

In this connection it is also notable that the reduction of $CO_2$ driven by a pH gradient has also been demonstrated in a microfluidic reactor system by Hudson et al [136].

One important point not yet addressed is how organizational complexity itself evolves. This is taken up next.

## 12 Contingent incrementalism and the ontogeny of complexity

As their name implies, Prigogine's 'Self Organizing Dissipative Structures', tornadoes for example, do not, nor could they, pop into existence fully formed. They instead form through an 'organizing' process – which begins with a single fluctuation-created 'seed' organization that functions as a conversion engine and which in consequence can grow incrementally and contingently to larger sizes. The humble 'dust devils' of common experience are a suggestive example. What is the nature of that process; what initiates it, what compels it to proceed forward, and what determines when it stops? Finally, is that process one in which the structure that is becoming progressively more organized, merely becoming larger, or is it adopting more complex organizational modes?

Our discussion of how order-converting engines organize matter tacitly assumed that the conversion engines themselves have fixed properties. This notwithstanding the fact that they themselves are organized, more-or-less complex, states of matter – and cannot have come into existence or have been maintained by any path other than that responsible for all the rest of living matter's ordered states. And of course, even though a living system's aggregate organizational complexity is necessarily maintained by its order-converting engines, they are themselves a small part of that system-level complexity. Yet neither they, nor any of the rest of life's organizational complexity, could have come into existence as the extremely complex organizations of matter they now are. So under what forcing mechanisms and what constraints, and by what sort of path, does complexity itself increase in living systems? Or for that matter abiotically? In particular, since we are now compelled to view organized states through Boltzmann's fluctuation-colored eyes, is making an order-converting system more organized at the same time making it more efficient at "sucking in" and converting "orderliness" – and can that fact itself be the thermodynamic force driving the evolution of that greater complexity?

Phenomenologically, the answer is a clear yes, as the mere, and commonplace, existence of SODS testifies. As we have already observed, such systems (e.g. vortex atmospheric phenomena) necessarily start small, as a local thermal fluctuation, and grow to sensible, or greater, size and power [15]. But these are mostly examples of growth in size, not growth in organizational complexity *per se*. We next consider the issue of dissipative engines evolving, on the fly, more complex organizational designs.

## 12.1 An abiotic example of incremental and contingent increases in organizational complexity: Taylor-Cuette flow

In Taylor-Cuette flow a fluid is introduced into an annulus between two vertical cylinders one or both of which can be made to rotate around the cylindrical axis. If they are, a drag force on the liquid close to the moving wall causes the liquid to move in a 'following' manner. At low and steady rotation speeds the liquid moves as a laminar, coherent whole. But at some speed the fluid motion transitions to a more complex pattern – at first, typically, into a stack of co-rotating rings which are themselves also rotating around their own ring axis. But as the speed is increased further, a point is reached where this pattern is replaced 'discontinuously' by an even more complex, finely structured one. Such pattern succession events can be repeated several times as the rotation speed is increased, producing a stair-case sequence of increasingly complex patterns. See the video watch?v=Fnz8uftHCEg.

In this sequence, each pattern can only arise as the contingent direct descendent of a less complex parent. Further, the transitions between the patterns are in effect discontinuous and occur at discrete 'fixed points' in the velocity variable. At those transition points the two patterns appear to be 'competing' on equal terms 'for the job', being presumably at that point equally efficient at dissipating the stress imposed by the rotation, and where the transition between competing structures seem clearly to be driven by fluctuation events. Above the transition point, the new pattern becomes increasingly stabilized as the established performance winner.

How does this phenomenology conform to theory? It is fair to say that this is a complex, not entirely settled area of contemporary research; but the general support for the qualitative picture outline above seems strong. See [137, 138] and perhaps particularly [139] and the discussion therein of "dissipative adaptation" buttressed by a remarkable 'macroscopic' example which the author describes in his penultimate section, "*The search for a general phenomenon*".

In summary, thermodynamically driven systems, we can reasonably suppose, are sometimes 'unstable' to transitioning into locally organized 'engine' states if those states accelerate, in the net, the rate at which the driving disequilibria is being dissipated – i.e., is generating entropy. And on the same thermodynamic logic, that system can be unstable to making 'discontinuous' further transitions to even more organized versions of themselves – if, again, those transitions 'pay off' in terms of further increasing the system's entropy production rate; net, that is, of the entropy consumption costs of the added organization.

But perhaps the key insight is that in general, the more organized forms cannot arise directly from the fully disorganized, equilibrium state. Instead, they must get there by paying their dues of passage; a passage comprising necessarily a staircase of incremental, thermal-fluctuation sized steps, each one requiring its predecessor to function, and being possible only in its context.

This conclusion then brings us to our final point about why chemistry, impoverished of engines, cannot breathe life into matter.

## 13 Dreams of transmutation; alchemy and bio-alchemy

No appropriately educated person would now contend that it might be possible, much less within the arts of 'chemistry', to transmutate base metals to gold. But imagine a nuclear physicist charged with the task of explaining to medieval alchemists, however brilliant and erudite they may be, why their primary goal is unattainable, their labors futile – and to an absolute certainty. How might she even begin?

The argument this piece has sought to advance analogously contends that the goal of transmutating matter from inanimate to animate states by any of the arts of chemistry is to an absolute certainty unattainable, and all effort to do so therefore futile. But indeed, for the most part the reasons are hardly less alien, "ridiculous", and unbelievable to almost all knowledgeable experts in the relevant life sciences – as, that is, would have been the principles of nuclear physics to the alchemists of history. This to no one's discredit. The relevant underlying realities have emerged late – in effect just yesterday, they lie well below the field's radar, and come from as as far out of the conceptual 'left field' of the science as could be imagined.

The villain in the story is, again, Boltzmann, or rather what his order-disorder principle implies when cast into the 'Chicago at rush hour' reality of the complexity of a living cell. That is due mainly to two facts. First, only a modest fraction of the improbability of state of a cell's organization is accounted for in the standing sum of all of the cell's endergonic processes. It is instead in the operational integration of them all into an inherently dynamic, self-steering, self-controlling, "fully autonomous, self driving" electrical vehicular system which, in the myriad of ways it must, 'works'. And which, to crown the matter, must also have historically organized itself, invented itself, bit by sequential bit, 'from scratch'.

But this is the issue treated just above, namely how 'complexity evolves', i.e., how systems functioning as 'engines of dissipation' initiate, and then become more complex. There we argued that they do so by what is necessarily an historical process; by ratcheting themselves stochastically up a staircase of very small increments of greater 'order' embodying small inventions, each one being entirely contingent on the context ('step') in which it arose, and adopted only if, in that context, it made the system as a whole better at 'sucking in orderliness' from the environment and thereby increasing the rate at which the system was disordering the environmental orderliness on which it fed – and doing so at a net gain-vs-cost benefit.

In developmental terms such a system is the paradigmatic 'shark'; not only can it never stop dissipating environmental orderliness, it cannot have gotten into its present state of organizational complexity, or any prior point in its evolutionary history of self organizing invention, except by having traversed the unbroken chain-link history of antecedents, of incremental and contingent organizational enhancements that brought it to that point – and just as at no point in that history can the flame of life have been everywhere extinguished without terminating it, so too at no point therein could it have been first ignited.

And this law of required, from the ground up, unbroken history dependence applies not just on the evolutionary scale. It applies as well to assembling the individual in its biological development. That is, an individual shark's functional organization cannot be brought into existence except by its development traversing, in the right sequence, an ontogenetic staircase of sequential and contingent 'next steps'. A staircase which itself can only have come into being as a part of an historical evolution of a phylogenetic staircase of incremental, 'next step' inventions. Where ascending either staircase requires starting at the bottom and staying on it. And where neither the ontogenetic nor the phylogenetic history – to a certainty – can be started up 'in the middle'. 'Frankenstein', as potent a metaphorical allegory as it is, utterly and 'fatally' misses what life is; see in particular the paper by Gary Sawyers [140] [21].

---

[21] A recent Perspective by Gary Sawers, *Perspective elucidating the physiology of a microbial cell: Neidhardt's Holy Grail* [140] incitefully explores in both organismal and thermodynamic terms, these and related points. Specifically those regarding the inescapable continuity and historically contingent developmental dependence of the organizational complexity of living systems. In this the key point is made that the required organization of the living system's matter simply cannot 'spring into being' (not even with all potentially needed 'engines' added); the system and its historical antecedents are mechanistically and causally inseparable. And are so, at root, because of Boltzmann's "order-disorder"

## 14   Conclusions

Arguably the most profound and 'overturning' insight to emerge from Boltzmann's fluctuation-based refounding of thermodynamics is what it ultimately revealed about what can and what cannot cause a physical system to become organized. To move, that is, counter-entropically away from equilibrium into an ordered, thermodynamically improbable, reduced-entropy state. Exactly the type of state, we note again, in which all living matter must necessarily exist.

But, as this piece has labored to explain, what that insight has revealed about how matter can come to be in such a state is not only highly counter-intuitive, it is in stark conflict with accepted views, both lay and professional. Views which in particular conceptually underpin nearly all thinking on the emergence of life. It would seem, therefore, that this conflict cannot conscionably be ignored given the obvious fact that the matter comprising a living system is – and as a necessary condition of being 'alive' – in a very specific, very far from equilibrium, organized state.

So, what do we know for a thermodynamic certainty about the ways the system does not - could not - come to be in such an elevated state?

The lead fact is that it is not, contra the consensus belief on the matter, by 'adding free energy', of any type or in any manner whatever, to some 'chemistry'. Not only can doing so not, as a matter of thermodynamic law, cause the system to become more organized; it would inherently make matters worse; a point eloquently made in the Benner *et al* paper discussed above [39]. Namely that any 'chemistry', fed free energy of any type, or in any manner, is compelled to "devolve" toward a kind of maximally disorganized 'asphaltic' death.

On the other hand virtually every element of the thermodynamically correct answer as to how matter can wrest itself from equilibrium's quick-sand in order to become organized, is both alien and heterodox to the seeming point of being palpably ridiculous. To review:

1. a system can be transitioned macroscopically into states of greater order only by the serial accumulation of incremental transitions aligned along the same dimension of order, each such transition resulting from an anti-entropic thermal fluctuation occurring entirely by chance.

2. the anti-entropic fluctuation transitions must be individually paid for thermodynamically in real time as they occur; but in principle the only way this can happen is by having each such occurrence trigger (i.e. permit) a statistically stronger, pro-entropic fluctuation in a different, but pre-ordered, 'driver' system.

3. in this way the dissipation of an increment of order in a driver system (involving, that is, an incremental *disordering* of that system), can be used to 'trap', after the fact, and render usable, the creation of an increment of order in a 'driven' system; the law is: only by order dissipated can order be created – and this can be made to happen only via 'trades', or 'exchanges' of order which are consummated one pair of (thermodynamically opposed) thermal fluctuation increments at a time.

4. a 'ratchet' mechanism must be provided for the sequential and more-or-less irreversible accumulation of the individual increments of increased order into a macroscopic extent of order created in the driven system.

5. the above operations can only be carried out by a 'be-spoke' macromolecular device acting technically as an engine; one whose fuel is order in one system and

---

insight – which Sawers also invokes using Schrödinger's framing and terminology; regarding specifically the fact that only by the mechanistically contingent and incremental dissipation of "orderliness" in one system can it be created in another.

whose work product is order in another; an engine that is, carrying out individual fluctuation-level order-for-order exchanges; ones that are specific as to both the states being disorganized and those being organized.

With respect to speculations about the emergence of life the central implication of these facts is both sharply restrictive and proscriptive. Models which do not countenance the need for the required 'engines', much less provide at least in principle a way for them to arise naturally (initially abiotically of course) and be properly fueled and employed, are inherently incapable of 'organizing' matter, (much less of escaping 'Benner's curse'), and are thus not thermodynamically defensible ideas.

We emphasize, however that the arguments advanced in this piece in no way discard all thinking on the origin of life; only that which would, even if only implicitly, require the violation of thermodynamic law. Where, again, the violation at issue is simply in regard to how matter can become organized into non-equilibrium states.

# Acknowledgments

The author owes a particularly essential debt to Dr. Michael Russell: for having set him on the path in the first place, and guided, informed, mentored, and encouraged his work on the problem thereafter – that represented in the present manuscript not least. And in like measure and for similar reasons he also owes a great debt to Dr. Wolfgang Nitschke, a kinder tutor to a befuddled 'newbie' could not be found. The author also thanks the Carl R. Woese, Institute for Genomic Biology at the University of Illinois for having extended to him the privileges of an Affiliate Faculty position.

# Appendixes

## A  The thermodynamic life of a single biological macromolecule doing work

As we have already indicated, the metabolic processes that do work – either physical (e.g. kinesin) or chemical (e.g. Na-K ATPase) – are not examples of an enzymatically catalyzed chemical reaction; they are instead instances of something much more intricate and foundational: the "coupling" of two processes by a macromolecule that forces the processes to precede under two constraints. The first is that they must move *in thermodynamic opposition*: one, the 'driven' process, moving 'in reverse', i.e. anti-entropically ("endergonically") and doing the work, the other a 'driving' process providing the 'power' and moving in its 'forward', pro-entropic ("exergonic") direction in order to do so [22]. The second is that 'instances' of their passing within the macromolecule are made to *reciprocally gate* each other whereby an instance of the process moving anti-entropically is used to 'gate' an instance of the process proceeding 'pro-entropically', and is discarded unless it manages to do so before it itself reverses.

To force two processes to interact in this contrarian 'pushmi-pullyu', inter-meshed, non-equilibrium manner, they must be under the control of a macromolecule (or complex thereof) functioning as a 'free energy converter', i.e. as an 'engine' [23]. All metabolic processes doing work are instances of the 'anti-entropic' member of such a coupled pair. And, notwithstanding essentially all accepted views on the issue, the exergonic process does not drive or power the endergonic one by transferring energy to it. In fact, how such coupling actually works is the crux of the questions being addressed in the present paper. In any case, it should already be clear from the above that such coupled, work-performing processes do not fall within the reach of conventional chemical thermodynamics. But there is much more to note in this regard:

1. Such conversion engines necessarily operate by cycling through an ordered and contingent sequence of distinct states. In each engine cycle, an appropriately 'metered' increment of each of the two coupled processes takes place; typically just one to a few copies of one molecular event (e.g. in ATPsynthase, each cycle carries out the synthesis and strongly endergonic export of a single ATP molecule from one ADP and one Pi, coupled to the exergonic, down gradient, passage of $\sim 4$ protons across a mitochondrial membrane).

2. However, for this to work productively the engine must impose a 'neither unless both' rule, according to which a cycle is not allowed to complete unless within it, one metered increment of each of the two paired processes has taken place.

3. Where we emphasize again that the anti-entropic member of the pair, i.e. the one 'violating' the classical 2nd law of thermodynamics – (though, of course, not the true, 'post Boltzmann' fluctuation-based 2nd law) is the one producing the engine's work output (equivalently, producing 'free energy').

4. In general such engines couple processes operating far-from-equilibrium (often extremely far), and carry out inherently dynamical processes to do so. So the objects of analytic interest here are **not** those of classical thermodynamics, i.e. quasi-equilibrium states; they are instead 'trajectories' comprising a stochastic sequence of dynamic, non-equilibrium transitions between different quasi-equilibrium and meta-stable macrostates of the system.

5. A conversion engine necessarily creates more entropy (in its exergonic process) than it destroys (in its endergonic one). And although it need not do so in every cycle, the thermodynamic forces that make the engine cycle at all see to it automatically that, more often than not, it executes cycles in the direction in which that cycle individually contributes more entropy than it consumes.

---

[22]That is, as we have emphasized, such coupling simply does not work by the transfer of energy between the coupled processes; a fact known and firmly established in the theory of such processes for half a century [75, 141] and foreseen on general grounds by Boltzmann and then later by Shrödinger long before that.

[23]Such engines are not catalysts; which, of course, can only accelerate processes that can take place in the absence of the catalyst.

The upshot is that a thermodynamic analysis of coupled (i.e. free energy converting) processes must be able to treat a single macromolecule executing a truly dynamical *process*, even one operating extremely far from equilibrium – not just quasi-equilibrium states in systems being analyzed in the 'large particle number' limit. In the last two to three decades a theory up to this task has emerged – the key elements of which are sketched next.

Consider a single macromolecule in solution along with small 'reagent' molecules with which it can interact – either by binding (or unbinding) one at a specific binding site, or by causing a bound molecule to undergo a catalyzed reaction, or by undergoing a conformational or orientation change in response to such an event. The central insight in approaching this problem is that this system's thermodynamic "degrees of freedom" resolve into two sharply distinct classes: 'mechanical/thermal' ("M/T") degrees of freedom comprising the position and momentum of the system's constituent atoms, and 'chemical/molecular' ("C/M") degrees of freedom specifying which reagent molecules are bound to the macromolecule and which conformational state or context it is in. Somewhat ironically, whereas the first of these defines its own, 'classical', equilibrium thermodynamic system; the second adds on top of that a system operating under the radically non-classical principles of non-equilibrium, individual macromolecule, trajectory thermodynamics.

Two key factors separate these classes in our single macromolecule system. First, its M/T degrees of freedom are sufficiently great in number and, when perturbed, sufficiently quick to reestablish thermal equilibrium [80, 146] [24], that much of the system's time is spent in one or another well-defined M/T equilibrium state – waiting for its 'Brownian' chance to move, either forward to its next state, or backward to its previous one. Where here equilibrium means that the system's internal energy states are distributed according to a Boltzmann-Gibbs ("BG") distribution with well-defined values for the internal energy's expectation value, its entropy, and its Helmholtz free energy (conventionally labeled, respectively, $\bar{E}, S, F$; see Seifert [14]). Second, relative to the speed with which the M/T degrees of freedom equilibrate, the changes of state involving the C/M degrees of freedom: binding/unbinding events, catalysis, conformational changes, take place on a much longer time scale; typically between a few hundreds of a microsecond, and a few tens of a millisecond [147–149]; e.g. ATPsynthase typically runs at a bit more than 100/sec, being fed at the same rate by a like number of electrons moving down an ETC [98].

However, when they do occur (e.g. a reagent molecule binds), the M/T degrees of freedom are forced to relax into a new MB distribution (i.e. with new values for the system's, average internal energy, entropy, and Helmholtz free energy).

Importantly, the new C/M state will in general have significantly different propensities with respect to subsequent C/M events (different binding affinities, catalysis rates, conformations, orientations with respect to external structures (e.g. membranes, microtubules). So the dynamics of such systems comprises a series of transitions in its C/M degrees of freedom each putting the system into a new state which is: semi-stable in terms of those degrees of freedom, and in a new BG distribution with respect to its M/T degrees of freedom – reflecting that potentially the macromolecule now has different behavioral predispositions with respect to a subsequent C/M transition. And biology, it is fair to say, exploits this potential to the hilt: 'designing', for each separate system, in what order each of these C/M transitions is to occur, and what new C/M properties will be in play after each transition – in order to achieve the job at hand. It is here that stochasticity takes center stage in the story; that is because both when the next transition takes place, and whether it is a forward or backward move, is a random choice, – not a deterministically defined one. With that in mind we turn to the specific case of systems converting free energy.

As has been noted, in such conversions the macromolecule is functioning as an engine, 'cycling' through an ordered sequence of states each of which has distinct C/M properties. In this cycle, as William Jencks emphasized [82], the two coupled processes are typically managed as 'half reactions' which are interleaved in a particular alternating order: endergonic-first-half, exergonic-first-half, endergonic-second-half, exergonic-second-half – where each step is contingent on the success of its predecessor. Phrased somewhat more concretely and in the terminology of this discussion, the conversion engine typically starts a cycle in a C/M state in which it can only facilitate one specific C/M event, i.e. binding a specific reagent molecule (or molecules), typically those comprising the 'products' in the reaction that the engine is set up to drive in reverse. When that binding occurs, the engine switches to a C/M state in which it can now only facilitate the binding of the reactants for the reaction that is to proceed 'forward' (i.e. exergonically). If that also happens, the engine has then completed the first 'half reactions' of the two processes – and it then switches to a conformation in which it can catalyze the endergonic reaction and then release that reaction's products –

---

[24] In this Brownian thermal fluctuation world, as Astumian has pointed out [12], the macromolecule is buffeted by water molecules $\sim 10^{21}/sec$.

thereby completing the entropy reducing, work-performing part of the conversion process [25].

Then the system faces its crux challenge for that cycle in the form of a kinetic competition: the just-created reduced-entropy state inherently has a relatively high probability rate for reversing; that is, an endergonic fluctuation is more or less strongly biased in favor of quickly being undone – and thus contributing to no useful work process. But, of course the engine, as much as it can be, is ready for that. It has its 'exergonic' barrel already loaded (e.g. with ATP) – and a finger 'twitching on its trigger' – set to detect the moment the endergonic process has completed. Thus the engine secures a chance to win the competition and save the precious endergonic fluctuation if: 1) it responds to detecting that uphill event by firing off the loaded instance of the exergonic process before the endergonic fluctuation reverses; and 2) if that exergonic process is irreversible enough such that the free energy lost when it takes place is greater than the free energy gained in the endergonic transition it is seeking to 'trap', then the exergonic reaction would have satisfied not just the kinetic 'race' requirement of the trapping, but also its 'thermodynamic' one – that of insuring that the net entropy change of the two reactions together is positive. And if the engine sees to it also that the exergonic event would have to be reversed before the endergonic fluctuation could be (a rule the engines of biology generally stringently ensure), then, with the second 'half-reaction' of both processes completed, the engine is, and can safely be, triggered to change its C/M state once again, this time back to 'go', thereby harvesting the precious work-producing fluctuation and completing one cycle – ready to go again.

We emphasise here, because the point is so widely misunderstood, that the exergonic process in a free energy conversion serves *only* the function of preventing the reversal of individual, work-performing, fluctuations of its endergonic mate – thereby 'trapping' them; a function which has both a kinetic 'race' aspect, and a thermodynamic 'free energy balance' aspect. Wherein, of course, it is the serial accumulation of such trapped fluctuations, one for each engine cycle, that yields the conversion process's macroscopic, order-creating *raison d'être* product. **To which essential function the exergonic partner contributes only sufficient (and controllable) irreversibility. Not energy.**

Note too that each of the states of the cycle are instances of the quasi-stable BG states, that is, in equilibrium with respect to their M/T degrees of freedom. And also that in the cycle, the system is either in one or another of these states or in the process of transitioning between them. That is, the transitions between the macrostates of the cycle are 'rare' events compared with those that equilibrate the system in its M/T degrees of freedom within each C/M macrostate. And note also that a system that is actively converting free energy is doing so because at least one of its C/M degrees of freedom is more-or-less strongly out of equilibrium. And that it is the inherent tendency of systems in non-equilibrium states to 'cool' when they can; i.e. to move progressively toward states closer to equilibrium; it is this (statistical) tendency which moves the engine through the steps of its cycle.

So: such systems are in tight equilibrium with respect to their 'fast' thermal/mechanical degrees of freedom while being at the same time more or less far out of equilibrium with respect to their 'slow' chemical/molecular degrees of freedom. But, as just noted, it is these disequilibria which cause engines whose cycling will dissipate them to do so. But, in this the macromolecule mediating the process is not just acting as a fatter pipe, allowing the dissipating fluxes to be greater than were the macromolecule not engaged. It is a device putting that dissipation to use. Specifically, in free energy converting engines the dissipation of disequilibria in one system's C/M degrees of freedom is, somewhat paradoxically, put to the good employment of doing work, i.e. of creating a disequilibrium in another system's C/M degrees of freedom (e.g. the electrochemical disequilibrium driving electrons 'down' the 'electron transport chain' in mitochondria, being converted to the transmembrane electrochemical proton-denominated disequilibrium of the "proton motive force").

---

[25]Note that fluctuation-driven anti-entropic 'half-reactions' are an essential part of even conventional chemistry; e.g. in a conventional chemical reaction both forward (pro-entropic) and reverse (anti-entropic) events must be taking place, each making its essential contribution to the Gibbs free energy for the reaction. But further that both of these reaction events require anti-entropic fluctuations to elevate either reactants or products to the transition state.

Which brings us to this aside's final point. We have noted that conversion engines operate by moving through a cyclic ordered sequence of discrete M/T-equilibrated stations in a "stochastic" process – behaving therefore as radically non-classical, non-equilibrium, single macromolecule machines governed by the principles of "trajectory thermodynamics". But what sort of process is that? In brief, it is a physical system 'hopping' along a discrete, ordered, circular sequence of distinct C/M states (each one, as we have noted, an equilibrated 'BG' state with respect to its M/T degrees of freedom) – wherein each hopping transition takes place between neighboring states in the sequence (in ideal operation). However, neither when a hop happens, nor whether it moves forward or backward in the sequence is deterministically constrained. Rather both of these outcomes are governed by probability distributions: a probability rate (i.e. the probability that the transition will occur per unit time), and separately a conditional probability that given the transition it will move forward rather than backward in the sequence; where both depend only on the current state, all memory of how the system got there having been lost in the M/T equilibration of the state. In formal terms, that is, these are Markov processes. The theory of such processes must then enable the probability of each possible sequence of moves (including all possible times and directions) along the sequence to be rigorously computed. This is of course not a trivial problem and it is one that takes us conceptually even farther away from standard chemical reaction theory – along with its conventional thermodynamic assumptions. Fully into, we might celebrate, Paul Boyer's great 'binding change' insight. But it is the hand reality deals us as the sharply counter-intuitive and iconoclastic answer to one of the deepest questions in science: how is life powered?

## B  A model stochastic, fluctuation-driven engine converting the dissipation of the order in one system into the creation of order in another

Consider an aqueous volume partitioned into top and bottom halves by a membrane – and within which two types of particles are suspended: 'blues' and 'reds'. As a starting state both particle types are distributed such that they are more concentrated in the top partition than the bottom, but unequally so; red being more out of equilibrium than blue: e.g. reds are tenfold more concentrated in the top, blues only three times. In other words, both types of particles are in an 'improbable', reduced entropy, 'ordered' state; reds more so than blues.

The challenge is to put a device into the membrane which will automatically increase the concentration gradient in the blues (making them even more 'ordered') at the expense of decreasing (i.e. 'disordering') that of the reds. Of course, the device must necessarily mediate processes in which particles transit the membrane – but certainly cannot permit either type to just flow down-gradient independent of the other. This implies that there must be, within the device, controllable, type-specific channels through which particles can transit the membrane. Such transits, however, will necessarily take place as stochastic, thermal-fluctuation-driven single-particle events, necessarily occurring in both down- and up-gradient directions. So then the question is:

how can the device mediate the creation of increased order in, say, the less-disequilibrated blue system, at the expense of (i.e. by dissipating) order in the more disequilibrated red system? The answer is that on principle the device can do so if and only if it makes the down-gradient, entropy-increasing (exergonic) transit of a particle of one type, conditional on the 'coincident' up-gradient, entropy-decreasing (endergonic) transit of a particle (or potentially a small number) of the other type; an event which itself can occur only as a chance Brownian fluctuation. But how in principle can this be done consistent with the 2nd law? The answer has two parts. First, enforcing a pairing of individual events in the two processes, whereby the device only allows an individual down-gradient red channel event to happen in the interval between the occurrence of an up-gradient chance fluctuation event in the blue channel, and that event's spontaneous reversion (in biology this is generally achieved by having the device detect the blue up-gradient event and then trigger – in the sense of permitting) the red down-gradient fluctuation event. The second is that on average more entropy must be gained in the individual red transit than was lost in the blue one (as we've insured in assuming a greater strength for the red gradient). The key requirement of this device is then clearly that it must 'force' a down-gradient transit through the membrane of a red particle to be contingent on the 'coincidental' occurrence of an anti-gradient (bottom to top) transit of a blue particle. Taken literally, however, it in principle can do neither. Whereas fluctuations can be blocked, they cannot be bidden. So the device must wait for the anti-entropic blue fluctuation it is trying to catch, and then 'gate' the pro-entropic fluctuation which, if it is irreversible enough, can allow anti-entropic event to go forward 'legally' and be used by the device to trap and 'save' the order-producing anti-entropic event.

The following describes the operation of a minimal device that embodies these principles – and which therefore can create order from order:

1. The device sits in, and spans, the membrane.

2. Two channels, one capable of conducting individual red particles, the other individual blue particles pass through the device connecting the top and bottom partitions.

3. Each of these channels is terminated at both ends by 'gates' that the device can individually open or close; and each has within it a binding site specific for its cognate particle. We will call these gates LB, HB, LR, HR (for the low and high concentration ends of, respectively, the Blue and Red channels.

4. The device's starting state is with only one of its four doors open, that giving the bottom (i.e. low concentration) volume access to the blue channel. There it waits until a fluctuation causes a blue particle to enter the channel and bind to the cognate binding site within it.

5. The device then undergoes a 'conformational' change which closes the LB door and opens the HR door giving the red channel access to the high concentration volume. (In biology, these conformation changes are in general slow and rate-limiting because they involve protein moieties moving through viscous water in a critically overdamped motion.) Again, the device waits until a fluctuation causes a red particle to enter the red channel from the high concentration reservoir.

6. With both LB and HR states 'loaded', the device then responds by opening the HB door; where it again waits until the blue particle dissociates and drifts out into the 'high' reservoir – moving against, of course, a high probability rate for a blue particle undoing that exit.

7. But to minimize that loss of opportunity, the device undergoes another conformation change as quickly as possible which both opens the LR door and closes the HB door. And then again waits until the red particle dissociates and drift into the L reservoir. From which, critically, the risk of that departure being reversed is relatively low.

8. If that succeeds before the B channel is refilled from the H reservoir (such that the paired, opposed, chance events: the uphill LB-to-HB event and its downhill sister the HR-to-LR have gone to completion), the device detects this fact and responds by undergoing its final conformation change – back to the state in which both channels are empty and only the LB door is open.

Each such cycle of the model engine produces a one-particle increase in the order of the blue particles wherein the red particle system contributes only the relatively greater irreversibility of a one-particle relaxation of its ordered state. Again, no 'energy' changes hands; the machine must simply wait until the chance transits of the membrane barrier take place – led by the order-creating, anti-entropic fluctuation of a blue particle 'blundering' into a counter-gradient transit. And must, in the biological real world, respond to these chance events by executing conformational changes (which variously open and close the 4 doors of the two channels). But in the biological world, the device is (usually) a protein macromolecule and the needed conformational changes are notably time-consuming. With the consequence that in the real biological world, life is 'rate limited' by the speed with which its order-trading engines can operate. As we have noted, ATP synthase can only run a little over a hundred cycles per second – implying that a mature coli cell must be running approximately 40 thousand full OxPhos-ATPsynthase systems!).

## C  The cost of conversion speed, and the max power balance

Few would contest the assertion made earlier that the free energy of ATP is typically maintained physiologically at about 23 (in units of $k_B T$). Meaning, since the concentration of phosphate is essentially unchanged in the hydrolysis of ATP, that $e^{23} \approx 10^{10}$ is (at steady state) the factor by which the concentration ratio: $[ATP]/[ADP]$ is physiologically held above what it would be at equilibrium. And perhaps also concede that the free energy being carried about by ATP is proportional to the log of this factor; that is to say, proportional to the extent of the ATP/ADP disequilibrium; $\Delta G_{ATPhyd}/k_T \approx ln\left([ATP]/[ADP]\right)$. But the accepted view then hurries to note that the coupled reaction processes involving ATP, whether creating or using this ATP disequilibrium, are found to operate close to equilibrium. Specifically, as we noted above, that when this ATP free energy is used to drive an endergonic process such as ion-pumping, e.g. that mediated by $Na^+/K^+-ATPase$, it is observed that almost as much free energy is gained in the driven process as is lost in the driving one (here the hydrolysis of ATP). So the net loss of free energy in the coupling is modest. The bulk being it is thought "conserved", rather than dissipated as heat by being transferred to and driving, the driven process. This in conformity with the presumed 'fact' that coupling processes in general work by the transfer of the energy released in the driving process (e.g. by the breaking of Lippman's squiggle bond in the hydrolysis of ATP) being donated to the driven one. The compelled conclusion is then that the coupled pair of processes are operating as a unitary (albeit catalyzed) chemical process taking place very near equilibrium – often near enough to be readily reversible.

However, as we have argued, this view is based on a fundamental misunderstanding of how free energy conversion actually happens (see primarily Sec. 6.1); and also directly violates (post-Boltzmann) thermodynamic law. Moreover, even the general observation that in coupled reactions the net change in free energy is small and that therefore they can be fairly regarded as comprising a 'close to equilibrium' processes, is at best misleading, and at worst misses an important point as to how bioenergetics works; namely the question of what determines just how far from being an equilibrium process free energy conversion 'should' be made to operate? It is trivial to note, of course, that coupling processes cannot be running too close to equilibrium since when the two processes are in thermodynamic balance no net conversion, no net work, is done. But there is more to it than that. In fact, the degree of imbalance in a conversion process is itself of substantial quantitative importance because it effects two critical performance measures of the conversion: 1) the speed with which the conversion process operates; relatedly the power output of the engine; and, somewhat separately, 2) the degree to which the conversion engine's efficiency is compromised by waste cycles. The first of these points then raises the question of how far the entropy gain in the driving process should exceed the entropy loss in the driven one? In particular, is there a ratio of these two rates that maximizes the engines output power?

Under standard simplifying assumptions, and for conversions involving processes in which the fluxes are proportional to the forces – as is the case for all of those of life [62], and for heat engines as well, the answer to this maximum power question is remarkably simple: *the power output of the engine is maximal if the rate of entropy production by the driving process is, in magnitude, twice the rate of entropy reduction in the driven one.* See Appendix C.1.for the arguments supporting this assertion.

## C.1 The entropy production imbalance of a conversion engine operating at maximum power

It was noted above that an electron traversing the ETC of mitochondria experiences a gain in probability of $\approx 10^{19}$ fold. And also that this thermodynamic force drives the endergonic synthesis of, on average, approximately one ATP. "Endergonic" in that the process drives the [ATP]/[ADP] ratio out of equilibrium by $\approx 10^{10}$ fold – producing a proportionally improbable state of the ATP+ADP+Pi system. So this conversion, sitting at the pinnacle of the cell's bioenergetics, is hardly operating 'close to equilibrium'.

But these two probability change ratios have a surprising and revealing relationship; the probability gain factor of the electrons is close in magnitude to the square of the probability loss factor of the ATP/ADP mix); equivalently, in OxPhos (under physiological conditions) the entropy gained per electron is roughly twice that lost per ATP synthesized. This is not a coincidence. It exemplifies a very general law of nature; one indicating, as we will see, that the OxPhos system is designed to maximize the power (literally the work per unit time) that conversion produces.

That conclusion reflects that, on quite general theoretical grounds summarized next, the power output of a conversion engine operating in a linear flux-force regime is maximal if exactly half of the entropy gain experienced by the driving flux is converted to the entropy loss of the driven flux – to, that is, the flux of the 'order' the engine is generating as its work output. In other words, if the rate of entropy change in the driving process is in magnitude twice that of the driven process, the power produced by the engine will be maximal. But, this is equivalent to saying (invoking Boltzmann's 'order-disorder' law; Eqn. 1), that for maximum power the factor by which the probability of the driving flux changes must be the (inverse) of the square of that experienced by the driven flux. In the ensuing sections we present the formal derivation of this principle, first in the case of conversion engines involving chemical processes, and then in the case of heat engines [26].

**Chemical process conversions**

Recall that the standard free energy conversion theory begins with the relation for the rate of change of the entropy within a volume due to the presence therein of two 'irreversible' processes:

$$\dot{S} = \sum_{i=1,2} J_i X_i \geq 0 \tag{3}$$

where $J_i$ and $X_i$ are, respectively the flux (e.g. reaction rate), and the entropy change per unit of flux, termed the 'force', of the $i'th$ process; the former are necessarily positive, whereas one of the later can be negative (note the 2nd law stipulation that $\dot{S} \geq 0$). And where it is assumed that in general each flux depends on both forces. Thus, under the assumption of flux-force linearity we can write:

$$J_i = \sum_{k=1,2} L_{i,k} X_k \tag{4}$$

where the $L_{i,j}$ are necessarily positive 'conductances' which must also satisfy the Onsager reciprocal relations $L_{1,2} = L_{2,1}$. In the case that $E_1 > 0$; $X_2 < 0$; then given that the fluxes are such that $J_1 X_1 \geq J_2 |X_2|$ ('work in' not less than 'work out', so that $\dot{S} \geq 0$, this system would be reducing the entropy of process 2 at the expense of increasing (by a greater amount) that of process 1 – i.e. acting as a free energy converter.

The two fluxes can thus be written as

$$\begin{aligned} J_1 &= L_{1,1} X_1 + L_{1,2} X_2 \\ J_2 &= L_{2,1} X_1 + L_{2,2} X_2 \end{aligned} \tag{5}$$

But in the simple leak-less conversion engine model considered here, both of these fluxes equal the flux through the cycle, $J = J_1 = J_2$ for all legal values of the forces. This requires that the conductances $L_{i,j}$ are all equal; here taken to be $L$, so that also the flux through the cycle is,

$$J = L(X_1 + X_2) = L X_{eff} \tag{6}$$

where $X_{eff} = X_1 - |X_2|$, is the 'effective', or 'net' force driving the engine's cycle. Note that this expression states that the engine's flux rate for converting free energy is proportional to the (net) force driving that flux.

Therefore, since the power out $P_{out} = -J_2 X_2 = -L \left( X_1 X_2 + X_1^2 \right)$, its maximum value is at the point at which $\partial P_{out} / \partial X_2 = 0 = -L \left( X_1 + 2 X_2 \right)$. Implying that:

$$X_1 = -2 X_2 = 2|X_2| \quad \text{at max power} \tag{7}$$

Thus in this model engine, at maximum power out the rate at which entropy is gained in the driving process is equal in magnitude to twice the rate at which entropy is reduced in the driven one: $\dot{S}_{driving} = J_1 E_1 = -2 J_2 E_2 = 2 J_2 |E_2| = -2 \dot{S}_{driven}$ for a conversion efficiency at maximum power of 50%.

---

[26] The following section is essentially a copy of Appendix E in our earlier paper [90], reproduced here for convenience.

**Heat engines**

The same principle as to the ratio of entropy-production rates between driver and driven processes in a conversion operating at maximum power applies also to heat engines [90, App.E]. There, however, the two fluxes being coupled are two fractions of a single flow of heat $J_{in} = \dot{Q}_{in}$ entering the engine at $T_{hot}$ – and delivering entropy into it at the rate of $\dot{S}^{in} = J_{in}/T_{hot}$.

The flow of heat (and of the entropy carried by it) into the engine is driven and determined by a temperature 'half-gradient', specifically that part of the total gradient between $T_{hot}$ and an intermediate "conversion temperature" here labelled $T_{conv}$, $T_{hot} > T_{conv} > T_{cold}$ defined as the point, in the total gradient, at which a fraction of the down gradient flow is converted to work. Therefore, the heat flux into the engine can be expressed as $J_{in} = L(T_{hot} - T_{conv})$ where $L$ is the thermal conductance of the path between the two indicated temperatures. So the corresponding rate of entropy import is: $\dot{S}^{in} = J_{in}/T_h = L(T_{hot} - T_{conv})/T_{hot} = L(1 - T_{conv}/T_{hot})$.

At the conversion temperature the flow of heat is interdicted, and a fraction of the entropy flowing into that point is converted to work (specifically into the rate of doing work; i.e. into 'power')[27] and the remainder carried away in the lower half gradient (between $T_{conv}$ and $T_{cold}$) where it is exhausted from the engine at temperature $T_{cold}$[28]. Of course this conversion is conventionally treated in energy rather than entropy terms; in which the incoming heat energy flux is divided into the part converted to work and the remainder exhausted from the engine: $J_{in} = J_{work} + J_{exh}$; subject to the condition that the entropy of the exhaust flow leaving the conversion point must not be less than that the entropy flow entering the engine: $\dot{S}^{exh}_{T_{conv}} = J_{exh}/T_{conv} \geq \dot{S}^{in}_{T_{hot}}$. Furthermore, the efficiency of this conversion will be the Carnot value $\eta^{Carnot} = 1 - T_{conv}/T_{hot}$ for this half gradient, providing that the rate at which the exhaust entropy is being removed is not greater than, but is exactly equal to, the rate at which entropy entered the engine.

However $J_{exh}$ is determined by the strength of the thermal conduction in the lower half gradient: $J_{exh} = L(T_{conv} - T_{cold}$ (assuming that the top and bottom half-gradients have equal thermal conductances) so the rate at which the entropy is removed from the conversion point must equal the rate at which entropy entered the engine: $S^{exh}_{T_{conv}} = J_{exh}/T_{conv} = \dot{S}^{in}_{T_{hot}}$.

*Thus, to allow the conversion to be taking place at maximum power the two half-gradients must be producing entropy at exactly the same rate. Implying, since all of the entropy production in the top gradient is converted to work, that the total entropy production of the gradient must be in magnitude exactly twice that 'lost' in the conversion to work.*

This in turn implies that the value of the conversion temperature must be such that $T_{conv}/T_{hot} = T_{cold}/T_{conv}$, or in other words that $T_{conv} = \sqrt{T_{cold}T_{hot}}$ so that the actual conversion efficiency at maximum power is:

$$\eta^{max\,power} = 1 - \sqrt{\frac{T_c}{T_h}} \tag{8}$$

And we have shown that just as for the 'chemical' free energy conversions considered above, at maximum power the entropy lost through being converted to work by a heat engine, is equal in magnitude to twice the entropy gain produced by the driving gradient.

---

[27] The conversion itself, here treated as happening in a 'black box', is assumed to take place in full conformity with the general Brownian/Boltzmann fluctuation principles discussed in the main text.

[28] Note that in the conventional 'Carnot' analysis of heat engines it is assumed that $T_{conv} = T_{cold}$.